\documentclass[acmsmall,nonacm]{acmart}
\AtBeginDocument{%
  }

\usepackage{xspace}
\usepackage{subcaption}
\usepackage{xcolor}
\usepackage{multirow}
\usepackage{todonotes}
\usepackage{booktabs}
\usepackage{pifont}

\usepackage{algorithm}
\usepackage{algpseudocode}
\usepackage{array}
\usepackage{textcomp}
\usepackage{stfloats}
\usepackage{verbatim}
\usepackage{graphicx}

\newcommand{\sysname}{S{\footnotesize warm}F{\footnotesize uzz}\xspace}
\newcommand{\sysOld}{\textcolor{black}{\sysname{G{\footnotesize raph}}}\xspace}
\newcommand{\sysNew}{\textcolor{black}{\sysname{B{\footnotesize inary}}}\xspace}

\newcommand{\graphname}{S{\footnotesize VG}\xspace}
\newcommand{\atkname}{SPVs\xspace}

\DeclareMathOperator*{\argmax}{\arg\!\max}
\newcommand{\new}[1]{\textcolor{black}{#1}}
\newcommand{\cmark}{\textcolor{green!80!black}{\ding{51}}}
\newcommand{\xmark}{\textcolor{red}{\ding{55}}}
\newcommand{\newtcps}[1]{\textcolor{black}{#1}}

\begin{document}

\title{Framework for Discovering GPS Spoofing Attacks in Drone Swarms}

\author{Yingao~(Elaine)~Yao}
\authornote{Work done when Yingao was a master student at the University of British Columbia.}
\affiliation{%
  \institution{Cornell University}
  \city{Ithaca}
  \country{United States}}
\email{elainey@cs.cornell.edu}

\author{Pritam~Dash}
\affiliation{%
  \institution{University of British Columbia}
  \city{Vancouver}
  \country{Canada}}
\email{pdash@ece.ubc.ca}

\author{Karthik~Pattabiraman}
\affiliation{%
  \institution{University of British Columbia}
  \city{Vancouver}
  \country{Canada}}
\email{karthikp@ece.ubc.ca}


\begin{abstract}
 Swarm robotics, particularly drone swarms, are used in various safety-critical tasks. While a lot of attention has been given to improving swarm control algorithms for improved intelligence, the security implications of various design choices in swarm control  algorithms have not been studied. We highlight how an attacker can exploit the vulnerabilities in swarm control algorithms to disrupt drone swarms. Specifically, we show that the attacker can target a swarm member (target drone) through GPS spoofing attacks, and indirectly cause other swarm members (victim drones) to veer from their course, resulting in collisions. We call these Swarm Propagation Vulnerabilities (SPVs). 

 In this paper, we introduce two fuzzing tools, \sysOld and \sysNew, to efficiently find SPVs in swarm control algorithms. \sysOld uses a combination of graph theory and gradient-guided optimization to find SPVs. Our evaluation on a popular swarm control algorithm shows that \sysOld achieves an average success rate of $48.8\%$ in finding SPVs. However, \sysOld fails to find any SPVs in drone swarms with different topologies. We then propose \sysNew, which uses observation-based seed scheduling and binary search to find SPVs. The evaluation shows that \sysNew’s success rate is comparable to \sysOld and work in all tested algorithms.
\end{abstract}




\maketitle

\section{Introduction}
Drone swarms are a type of distributed cyber-physical system that consists of multiple drones that communicate with each other. 
Drone swarms rely on swarm intelligence~\cite{swarm_def} to collaboratively accomplish the mission.
They carry out large-scale missions that cannot be performed by a single drone, e.g.,
logistics, surveillance, search and rescue~\cite{application1, application2, application3}.

Unfortunately, drone swarms are vulnerable to threats such as logic flaws \cite{swarmflawfinder} in swarm control algorithms and masquerade attacks~\cite{masquerade}.
However, attacks exploiting such threats incur high costs (e.g., introducing an external drone \cite{swarmflawfinder}, sending spurious messages to the swarm~\cite{masquerade}), and can be thwarted using techniques such as intruder detection \cite{intruder}
and drone authentication \cite{random-authentication}.
In contrast, physical attacks that feed the drone with erroneous sensor measurements via physical channels (e.g., GPS, accelerator, gyroscope) require relatively little effort, and can lead to drone crashes~\cite{gps-spoofing-1, gps-spoofing-2, acce-spoofing, gyro-spoofing, stealthy-attacks}. 

GPS attacks \cite{gps-spoofing-1, gps-spoofing-2}, 
which send malicious GPS signals to victim drones, are examples of physical attacks.  
GPS attacks have been shown in various systems such as self-driving cars, drones, and trucks \cite{gps-car,gps_drone,truck_gps}.
For example, during a drone show in Hong Kong, 46 drones crashed due to GPS jamming~\cite{gps-hk}. 
GPS spoofing, in which the GPS measurements are falsified, can be more debilitating than GPS jamming as it uses more subtle signals, and thus is more difficult to detect~\cite{spoofing_jamming}.  

In this paper, we first demonstrate that GPS spoofing attacks can disrupt drone swarms by exploiting vulnerabilities in swarm control algorithms  - we call these Swarm Propagation Vulnerabilities (\atkname). 
Specifically, the attacker launches GPS spoofing in a swarm member, referred to as the target drone, causing the target drone to deviate from its intended trajectory while avoiding collision with other swarm members, thus maintaining the stealthiness of the attack. 
The deviation changes the inter-distance between the target drone and other members of the swarm, referred to as victim drones. 
Subsequently, this leads to incorrect control commands generated by the swarm control algorithm. 
These incorrect commands in turn, cause the victim drones to veer off course, resulting in collisions. 

Note that the target drone in the attack is {\em not} the one involved in the collision, and hence it is difficult to identify it as the ``bad apple". 
Also, current defenses \cite{pidpiper, semperfi} for GPS spoofing attacks in a single drone often ignore small GPS spoofing deviations (e.g., $0-10m$).
Even if such defenses are deployed in all swarm members, they will fail to detect this type of attack.
This allows the attacker to attack the other swarm members without being detected, causing collisions, thereby reducing the mission efficiency and leading to potentially catastrophic outcomes such as crashes. 

The main cause of \atkname are the design choices of swarm control algorithms.  
To generate the control commands (e.g., heading directions), the swarm control algorithm has to balance conflicting goals with different priorities. 
For example, for a swarm control algorithm to maintain the formation in the swarm, it may give higher weights to goals involving interactions among swarm members than those that avoid the obstacle. 
While most swarm control algorithms are tuned to balance these goals under normal conditions and avoid collisions, they are often unable to do so when the attacker can manipulate swarm members' perceived locations,  
such as through GPS spoofing attacks, at strategic times. 
\new{This may lead to collisions, which can occur between two drones or between a drone and an obstacle along its path. In this paper, we refer to the drones and the obstacle involved in the collision as on-path objects. 
}

To help defenders evaluate the resilience of the drone swarm mission against \atkname, 
we design a fuzzing technique that can capture the attackers' capabilities,
map them to the drone swarm, and proactively discover \atkname before running swarm missions.
Fuzzing is a testing technique that has been widely used to find vulnerabilities in real-world applications \cite{fuzz-1, fuzz-2, fuzz-3, fuzz-4, fuzz-5, swarmflawfinder}.
\newtcps{Therefore, we leverage fuzzing to systematically explore different ways that the drone swarm can be attacked.}
However, directly applying previous fuzzing approaches for finding \atkname in drone swarms is challenging for the following two reasons. 
(1) The input spaces are large for long missions and large drone swarms, making it difficult to efficiently find the target-victim drone pairs that are vulnerable to \atkname.
(2) They typically target failures due to the exploitation of memory vulnerabilities, while we target failures that cause collisions in the swarm via GPS spoofing attacks.



We make the following two observations about the attacker who wants to successfully exploit \atkname in drone swarms. 
(1) To maximize the attack impact with minimal effort, the attacker should find the most influential drone as the target drone and the drone closest to the \new{on-path objects} as the victim drone.
(2) To efficiently cause collisions, 
the attacker should find GPS spoofing parameters that can minimize the distance between the victim drone and the \new{on-path objects}.
We find that this is a convex optimization problem, which can be solved efficiently.

Motivated by the above observations, 
we propose \sysOld, a novel fuzzing technique to efficiently find \atkname in drone swarms.
\sysOld has two innovations. 
First, we develop an abstraction called the  \emph{Swarm Vulnerability Graph (\graphname)} \cite{seed_centrality} to measure the malicious influence of each target-drone pair. 
\sysOld leverages graph centrality \new{for seed scheduling, which is} to decide the order of the target-victim drone pairs for fuzzing, 
so that the most influential drone pairs are prioritized to discover the vulnerabilities more efficiently.
Second, \sysOld employs gradient descent, which is known to be efficient for convex optimization problems\cite{convex-optimization}, 
to search for the other GPS spoofing parameters that will cause collisions. \new{This process is called search-based fuzzing.} \newtcps{We originally described \sysOld in the conference version of this paper~\cite{swarmfuzz}}.


We apply \sysOld\footnote{Available at: https://github.com/DependableSystemsLab/SwarmFuzz} to \new{two} popular swarm control algorithms, namely Vicsek~\cite{Viscek} and Olfati-Saber~\cite{OlfatiSaber}, 
in the  SwarmLab\cite{swarmlab} simulator. We evaluate \sysOld in different swarm configurations by varying the swarm size and the GPS spoofing distance.
\new{We found that \sysOld achieves an average success rate of $48.8\%$ in finding \atkname in the Vicsek \cite{Viscek} algorithm. 
Furthermore, the \graphname boosts the success rate of \sysOld in finding \atkname by more than $10$x; and the gradient-guided optimization reduces the runtime of SwarmFuzz by $3$x compared to a random fuzzer. 
However, \sysOld did not find any SPVs in the Olfati-Saber algorithm\cite{OlfatiSaber}.}



\new{We identify the following two limitations in \sysOld that account for its inefficiency in the Olfati-Saber algorithm.
(1) \graphname lacks generality when dealing with swarms of varying topologies. This leads to inaccurate influence measurement of target-victim drone pairs in certain swarm configurations. 
(2) The gradient descent algorithm can overshoot the target unless the parameters are well-tuned, which requires a lot of manual tuning and is thus effort-intensive.
}

\new{To overcome these two limitations, we propose \sysNew, an improved version of \sysOld. 
\sysNew uses a topology-agnostic seed scheduling and parameter-agnostic search-based fuzzing.
Recall that in seed scheduling, we determine the target-victim drone pairs with malicious influence, known as the seeds for fuzzing.
\sysNew accomplishes this by observing how the victim drone responds when the target drone experiences varying amounts of GPS spoofing while navigating around an on-path object.
The underlying idea is that if spoofing a target drone for both short and extended durations causes the victim drone to avoid the object from different directions, then spoofing within this time range is likely to cause the collision.  
\sysNew then treats the target-victim drone pair involved in the above scenario as the seed for fuzzing. 
This overcomes limitation 1 as this is topology-agnostic. Then, in the search-based fuzzing, \sysNew employs binary search, instead of the gradient-based method, to find the GPS spoofing parameters resulting in \atkname. 
Unlike gradient-based method, binary search does not need parameters such as learning rate to adjust the search speed. 
This overcomes limitation 2 as binary search is parameter-agnostic.}

\new{We apply \sysNew to the same swarm control algorithms as \sysOld, with the same configurations. \sysNew's success rate is comparable to \sysOld in both the algorithms. \sysNew finds that, in different swarm control algorithms, \atkname could cause collision between the drone and the obstacle, collisions between two drones, or prevent a swarm member from making progress towards the destination.}

\new{\sysOld and \sysNew share a common framework involving seed scheduling and search-based fuzzing for finding \atkname in drone swarms. In seed scheduling, \sysOld employs a graph-based approach to measure the drone pair influence, yielding low runtime overhead but limited adaptability to different swarm topologies. In contrast, \sysNew takes a topology-agnostic approach by analyzing the victim drone's behavior when avoiding the on-path object, leading to more adaptability to various swarm topologies but higher runtime.
In the search-based fuzzing phase, \sysOld uses gradient-descent optimization on the convex collision function, requiring manual parameter tuning. 
In contrast, \sysNew transforms the convex function into a monotonic one and uses binary search, reducing manual effort (as it does not need parameter tuning) and enhancing efficiency.}

\new{Both \sysOld and \sysNew help swarm designers evaluate the resilience of the swarm mission beforehand.
If the swarm mission is found to be vulnerable to \atkname, 
the designers can take actions (e.g., tuning the parameters in the control algorithm) to make it secure.} 


{\itshape To the best of our knowledge, we are the first to demonstrate the presence of \atkname in drone swarm control algorithms, and propose a technique to systematically and efficiently find such vulnerabilities.}
We make the following contributions.

\begin{itemize}
	\item Demonstrate attacks that exploit  \atkname to indirectly cause disruptions (e.g., collisions) in drone swarms. 
	\item Propose the \graphname abstraction, and utilize centrality analysis to find drone pairs that are likely to result in collisions. 
	Design \sysOld, a fuzzer that uses \graphname and gradient-guided optimization to efficiently evaluate the resilience of swarm missions by finding \atkname that cause collisions. 
    \new{\item Design \sysNew, a fuzzer that uses the property of monotonic functions and binary search to efficiently evaluate the resilience of swarm missions by finding \atkname that cause collisions}
	\new{\item Evaluate \sysOld and \sysNew on two popular swarm control algorithms. 
	We find that (1) \sysOld can achieve an average success rate of $48.8\%$ in finding \atkname across swarm configurations, while \sysNew can achieve an average success rate of $60.2\%$ and $86.5\%$ in each swarm control algorithm;
	(2) \sysOld and \sysNew have a higher success rate for missions with a larger swarm size or with longer GPS spoofing distance. }
	\new{Further, the \graphname boosts the success rate of \sysOld by $>10$x;   
	and gradient-guided optimization reduces the overhead of \sysOld by $>3x$.
    Finally, the seed scheduling boosts the success rate of \sysNew by $>5x$; 
    and the binary search reduces the overhead of \sysNew by $>3x$.}
	

\end{itemize}

\section{Background and Threat Model}
In this section, we explain how drone swarms operate, followed by how GPS spoofing attacks are launched. Finally, we present the threat model. 
\subsection{Drone swarms.}
\label{II-A}
Swarm control algorithms can be either distributed or centralized~\cite{swarm_def}. Centralized algorithms are less resilient as they have a single point of failure. Therefore, we focus on distributed swarm control algorithms in this paper.

\begin{figure}[!ht]
	\begin{center}
		\includegraphics[width=0.44\textwidth]{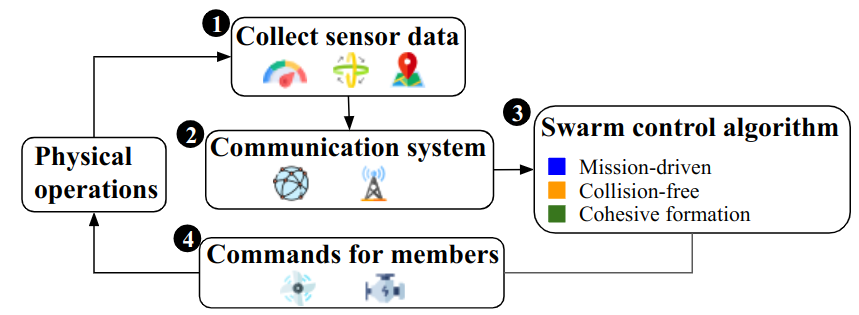}
	\end{center}
	\caption{Workflow of distributed drone swarm systems.}
	\label{fig:swarm_sys} 
\end{figure}

Distributed drone swarms follow four steps in a periodic loop ( Fig.~\ref{fig:swarm_sys}):
(1) each swarm member reads sensor data (e.g., GPS, IMU) to get its current physical states (e.g., location, velocity); 
(2) swarm members exchange physical states among themselves by sending messages through the communication system;  
(3) the swarm control algorithm running in each drone uses the physical states of other swarm members to compute the state difference (e.g., inter-distance, relative velocity); 
(4) each drone derives the control commands by itself independent of other drones based on the state difference and goals.

Specifically, to coordinate the swarm members, the swarm control algorithm has to adhere to the following three high-level goals:
(1) mission-driven, to ensure the drone swarm is  moving towards the destination; 
(2) collision-free, to ensure no collisions by maintaining minimum distance among drones/obstacles; 
(3) cohesive formation, to preserve the swarm formation by avoiding long distance among drones.

As a result, the final control command generated in a drone mainly consists of three sub-commands, each for a specific goal. 
One key component in the swarm control algorithm is to balance these potentially conflicting goals through careful tuning.
For example, when an obstacle is in the path of a drone to the destination, the swarm control algorithm needs to assign higher weights to goals  (1) and (2) to ensure that the drone moves  towards the destination while avoiding the obstacle.

\subsection{GPS Spoofing Attack}
The Global Position System (GPS) is widely used in Robotic Vehicles (RVs) for outdoor localization.
The GPS receiver calculates the position based on the
 signal received from the GPS satellites.
Civilian GPS systems lack signal authentication and encryption,
and hence are vulnerable to GPS spoofing \cite{gps-spoofing-2}.
RV missions like delivery, search, and rescue \cite{application1, application2, application3} are vulnerable to GPS spoofing as they depend on the location information. We focus on missions in which the swarm goes from one point to another.

In a GPS spoofing attack, the attacker transmits fabricated GPS signals with stronger power than the original GPS signal, so that the victim GPS receiver locks on to the attacker's signal.
The attacker controls the victim's perceived position by manipulating the GPS messages. 
GPS spoofing has been shown in laboratory  \cite{gps-spoofing-1} and real life settings \cite{gps-car, truck_gps, gps-hk}. 

Prior work has demonstrated how to infer a drone's information to perform GPS spoofing attacks on it~\cite{gps-spoofing-1, gps-spoofing-2, gps-spoofing-3, gps-spoofing-4, gps-spoofing-5}.
While many defenses against GPS attacks have been proposed for a single drone \cite{pidpiper, semperfi, gps-defense2}, 
most of them ignore small GPS  spoofing distances (e.g, $0-10m$) as those are indistinguishable from the  standard GPS offset \cite{gps-offset}, This is often sufficient to ensure the safety of a single drone.
However, in a drone swarm, 
since each swarm member acts based on its neighbors’ physical states,
 even a slight location deviation in one drone's location may influence the other members, and cause disruptions (e.g., collisions) in the drone swarm.

\subsection{Fuzzing}
\newtcps{Fuzzing is a software testing technique that has found many critical bugs and security vulnerabilities in real-world applications~\cite{libfuzzer, afl, syzkaller}. 
It has been adopted by major software companies such as Google~\cite{afl} and Microsoft~\cite{ms-fuzz} for security testing. 
Fuzzing starts with some random or specified inputs for the program under test, then performs the transformation on these inputs, and monitors the program for exceptions such as crashes. 
Since the input space for a program is huge to explore, most fuzzers use code coverage as feedback to measure the quality of the inputs produced by the fuzzer. This is based on the general observation that higher code coverage often leads to better crash detection. 
One of the leading coverage-guided fuzzing tools is American Fuzzy Lop (AFL)~\cite{afl}. 
However, since conventional fuzzers mainly focus on memory-related vulnerabilities such as buffer overflows, they cannot be applied to find SPVs in drone swarms.}

\subsection{Threat Model}

We assume the attacker can perform GPS spoofing attacks in only $one$ member in the drone swarm. 
\newtcps{To demonstrate the presence of SPVs in drone swarm control algorithms, it is sufficient to show that GPS spoofing attack on a single drone can expose SPVs. Such an attack would imply that similar attacks on multiple drones would also expose SPVs.}
This also requires much less attack effort than spoofing the GPS for multiple drones in the swarm, and makes the attack stealthier. 
\newtcps{
In practice, it is possible that the malicious GPS signal might slightly influence other swarm members if the attacker is not careful. For more precise control over the GPS signals, the attackers can take precautions such as reducing the transmission power of the GPS spoofer to the minimum, using attenuators to reduce the signal strength after the signal locking. For example, Zeng \emph{et al.}~\cite{gps-car} find that the GPS signal is too weak for drones two meters away to lock in with these efforts. If the drones are at least two meters apart, the GPS spoofing signals will not affect other drones.}

\new{The attacker's goal is to induce collisions within the swarm. These collisions can either occur between the drone and the obstacle collisions or between two drones.}

We assume the attacker knows the obstacle's GPS coordinates.
She can achieve this by either 1) placing the obstacle by herself, or 
2) choosing an existing obstacle in the drone swarm’s trajectory, and looking up its GPS location.
\newtcps{We assume that the attacker knows the drone swarm's trajectory. Since most missions are tested in a simulator before real-world deployment, the attacker can infer the trajectory by analyzing the simulation data. Other work has made similar assumptions~\cite{stealthy-attacks}. }
We also assume the attacker knows the swarm control algorithm being used, 
but she does not know the high-level goals explained in Section~\ref{sec:motivation} as \sysname automatically finds the vulnerability.

Further, the attacker does not have the capability to intercept or modify the messages exchanged among the swarm members as the messages may be encrypted \cite{Mavlink}. Finally, we assume that the attacker cannot introduce any external drones into the swarm. 

\section{Motivating example and Challenges}
\label{sec:motivation}
We first present an example drone swarm running a highly-cited swarm control algorithm - Vicsek algorithm \cite{Viscek}, to show how \atkname can be exploited (details in Section~\ref{sec:experiment}). 
We then present the design challenges in systematically finding \atkname{}.

\subsection{Motivating example}

\begin{figure}
	\begin{center}
		\includegraphics[width=0.44\textwidth]{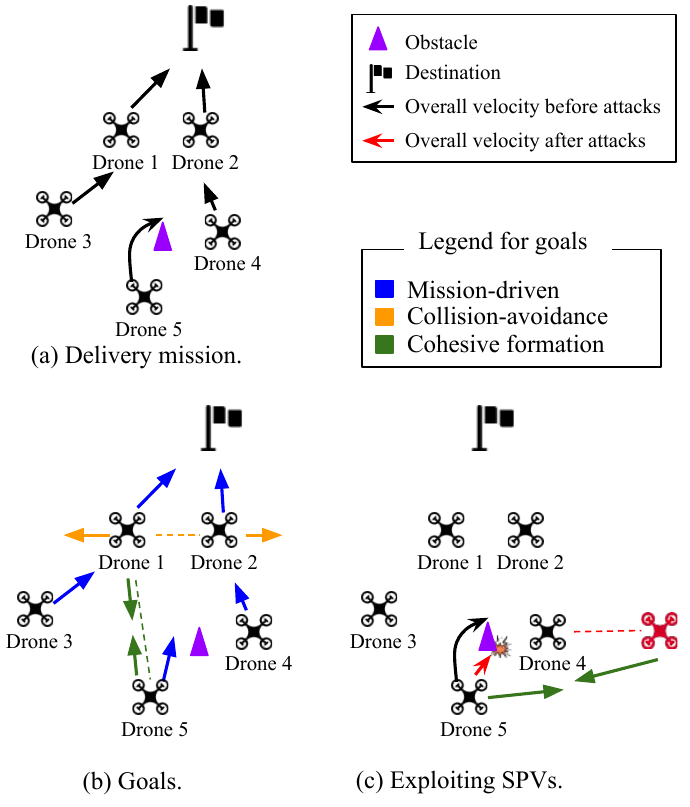}
	\end{center}
	\caption{Motivating example for \atkname in drone swarms. }
	\label{fig:motivation} 
\end{figure}

\textbf{Swarm Setup.}
We emulate a 5-drone swarm in the Swarmlab \cite{swarmlab} simulator for a delivery mission, as shown in Fig.~\ref{fig:motivation}-(a).
It aims to reach a pre-defined destination (i.e., the flag) while avoiding the on-path obstacle (i.e., the purple triangle).
For example, in Fig.~\ref{fig:motivation}-(a), the obstacle blocks drone 5's way to the destination.
Therefore, the swarm control algorithm attempts to make drone 5 avoid the obstacle from the right.

\textbf{Goals.}
As mentioned in Section~\ref{II-A}, the swarm control algorithm mainly follows three goals to generate the appropriate control commands, i.e., (1) mission-driven, (2) collision-avoidance, 
and (3) cohesive formation. The algorithm generates different sub-velocities for each swarm member, based on the above goals.
Fig.~\ref{fig:motivation}-(b) shows the mapping between each goal and the corresponding sub-velocity generated. 

For goal (1), each drone has a sub-velocity for moving towards the destination (i.e., blue arrows).
For goal (2), with the short distance between drone 1 and drone 2, repulsive sub-velocities (i.e., orange arrows) are generated to avoid collisions between them.
For goal (3), with the relatively long distance between drone 1 and drone 5, attractive sub-velocities (i.e., green arrows) are generated between the drones to maintain the formation.
These sub-velocities may conflict with each other. 

\textbf{Exploiting \atkname.}
The attacker first launches GPS spoofing and causes  the target drone to perceive a wrong location, which it communicates to the swarm. 
This deviation changes the inter-distance between the target drone and the victim drone,  
causing the swarm control algorithm to generate incorrect control commands.
These incorrect commands cause the victim to deviate from the course, 
potentially resulting in collisions.

For example, in Fig.~\ref{fig:motivation}-(c), drone 4 (target drone) is under the GPS spoofing attack and thus deviates to the right.  
This deviation increases the inter-distance between drone 4 and drone 5 (victim drone), making it difficult for the drone swarm to maintain the formation.
Hence, according to goal (3) in Section~\ref{II-A}, attractive sub-velocities (i.e., green arrows) between drone 4 and drone 5 are generated.
Note that originally, drone 5 avoids the obstacle from the right. However, with this new sub-velocity, the overall velocity (i.e., red arrow) for drone 5 points towards the obstacle, thereby leading to a collision. 

According to goal (2) in Section~\ref{II-A}, drone 5 also has a sub-velocity for avoiding the obstacle.
However, since the sub-velocities generated by other goals are bigger than the sub-velocity to avoid the obstacle, drone 5 flies towards the obstacle and collides with it, thereby violating the swarm's safety constraints. Thus, the above attack was successful. 

We found that the above attack succeeds in different swarm missions with different GPS spoofing distances (i.e., $5m$ and $10m$), 
indicating that this particular swarm configuration is highly susceptible to \atkname. However, it is tedious to perform this evaluation manually for different swarm configurations. 

\subsection{Design Challenges}

For defenders to evaluate the resilience of swarm configurations, 
it is important to develop an automated approach to systematically and efficiently find \atkname in drone swarms. 
This leads to two unique challenges, as follows.

\textbf{C1: Finding Target-Victim Pairs.} 
To exploit the \atkname{} and launch a stealthy attack, appropriate target-victim drone pairs need to be chosen strategically.
This is because in a large drone swarm, the number of possible combinations of target-victim drone pairs is huge, which leads to a large input space.

\textbf{C2: Choosing Spoofing Parameters.}
Given the spoofing deviation, performing GPS spoofing attacks involves two key parameters, i.e., spoofing direction and spoofing time.
Spoofing direction refers to the direction the target drone deviates under GPS spoofing (e.g., right, left), 
while spoofing time refers to the duration of time that the GPS spoofing attack lasts.
The wrong choice of the spoofing parameters (i.e., direction and time) in the target drone will 
cause the victim drone to avoid the collision, thereby defeating the point of the attack.  

\section{Methodology of \sysOld}
\label{sec:methdology}
\begin{figure}[!ht]
	\begin{center}
		\includegraphics[width=0.5\textwidth]{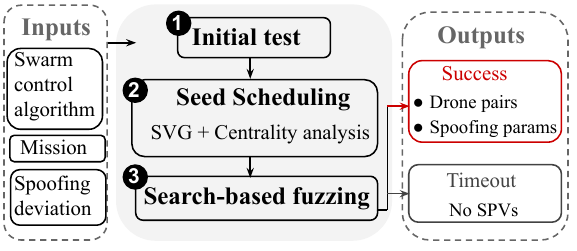}
	\end{center}
	\caption{Overview of \sysOld{}}
	\label{fig:overview} 
\end{figure}

Fig.~\ref{fig:overview} presents the overview of \sysOld{}. 
\sysOld{} only takes as inputs (1) the swarm control algorithm, (2) the mission parameters (including the swarm size and the location of the obstacle), and (3) the GPS spoofing deviation. 
It performs the following three steps. 
(1) It runs an initial test without any attack. 
If this test mission is successful (i.e., no collisions), it records mission information to construct the \emph{Swarm Vulnerability Graph (\graphname)}.
(2) It performs centrality analysis on the \graphname to analyze the influence of each drone with a certain spoofing direction, 
and uses this to decide the order in which target-victim drone pairs are selected for fuzzing (\textbf{C1}).
(3) For a certain target-victim drone pair with the spoofing direction, it searches for the spoofing parameters (i.e., start time and duration) that minimize the distance between the victim drone and the obstacle with gradient-guided optimization (\textbf{C2}). 
It repeats step (3) until a collision occurs or it reaches a predefined number of search iterations. 
If a collision occurs, \sysOld has found a successful SPV, and it outputs the  target-victim drone pair(s), and the spoofing parameters for the collision. Otherwise,   \sysOld reports that no \atkname were found in the mission, and the mission is resilient to SPVs.

\subsection{Test Definition and Initial Test Creation}
\label{IV_A}

A test-run is defined as a set of tuples $<T-V, t_{s}, \Delta t, \theta>$, 
where $T-V$ represents the target-victim drone pair,
$t_{s}$ represents the spoofing start time, 
$\Delta t$ represents the spoofing duration, 
and $\theta$ represents the spoofing direction. 
We aim for as few spoofing parameters as possible to minimize the attacker's effort, 
and thus focus on \emph{horizontal constant spoofing},
 i.e., always setting the GPS spoofing distance to a constant $d$ (provided as the input) during time $\Delta t$, and only perform spoofing horizontally.
Thus, the value for $\theta$ is +1/-1, representing right and left respectively.
The value range for $T$ and $V$ could be any drone in the swarm as long as they are different from each other.  
$t_{s}$ and $\Delta t$ range from 0 to the entire mission time.

We create the initial test case by running the swarm mission without any spoofing attack.
During the test, we collect the following  information:
(1) each drone $i$'s location ($x_{i}, y_{i}, z_{i}$) at each timestamp $t_{j}$, i.e., $<x_{i}, y_{i}, z_{i}, t_{j}>$;
(2) the minimum distance between each drone $i$ and the obstacle throughout the mission ${D_{ob}}^{i}$; and 
(3) the mission duration $t_{mission}$.

\subsection{Seed scheduling}
\label{IV.B}

To tackle \textbf{C1}, we measure the malicious influence of each target-victim drone pair, and 
choose drone pairs for fuzzing in decreasing order of the influence.
We first develop an abstraction called the \emph{Swarm Vulnerability Graph (\graphname)}.
We then utilize graph centrality analysis to measure the influence of a swarm member in the \graphname.
Finally, we choose the most influential member as the target drone
and the drone closest to the obstacle as the victim drone, to order the seeds for fuzzing.

\textbf{Seed pool.}
The seed pool consists of a set of seeds $<T-V, \theta>$.
We observe that the probability of causing collisions for each seed depends on two factors:
(1) the influence of the drone pair $T-V$;
and (2) \new{the Victim drone's closest Distance to the on-path Object (VDO) (in the absence of attacks). The on-path objects refer to the other swarm member and the on-path obstacle.}
For factor (1), the malicious impact of the target drone is a function of its influence over the whole drone swarm, especially the victim drone.
Thus, choosing the most influential drone as the target is more likely to cause collisions.
For factor (2), a drone closer to the obstacle (i.e. low VDO) is more promising as a victim drone, as it requires lower effort to crash into the obstacle. 

\textbf{\graphname Definition.}
The malicious influence of each drone can be intuitively measured by the swarm member's reactions to the spoofing deviation of the target drone.
To quantify this influence, we propose the \emph{Swarm Vulnerability Graph (\graphname{})}. 
The idea behind the \graphname{} is to utilize the centrality analysis in graph theory \cite{influence}, 
which measures a node's influence in the graph and is used to identify the graph's most influential node.

We define $SVG = (N,E,W)$, 
where $N$ is the set of nodes representing the swarm members, $E$ is the set of edges capturing the connections between drones, and $W$ is the set of weights measuring the local influence on each edge $e$.
\newtcps{The details on constructing the SVG can be found in Section IV-B in the original SwarmFuzz paper~\cite{swarmfuzz}. We do not present these here due to space constraints. }

\textbf{Centrality Analysis.}
The centrality of a node represents its influence. 
%
{\em PageRank} centrality ~\cite{pagerank} is efficient to compute for large graphs. Therefore, we choose {\em PageRank} to approximate a drone's malicious influence in the \graphname.


\textbf{Seed Scheduling.}
The probability of causing collisions for each seed depends on
(1) the influence score of the drone pair;
and (2) the VDO.
To schedule the seeds based on these two factors,  we
(1) sort each victim drone based on the VDO in the ascending order;
(2) calculate the summative influence of each possible combination of drone pairs; 
(3) for each victim drone, choose the target drone involved in the drone pairs that have the highest summative influence.

\subsection{Search-based fuzzing}
\label{sec: convex}
To tackle C2, our goal is to find the appropriate spoofing parameters (i.e., $t_{s}$ and $\Delta t$) that will cause the collision, given a specific seed. 
We model this as an optimization problem - the goal is minimizing the objective function $f(t_{s}, \Delta t)$, the distance between the victim drone and the obstacle, 
subject to the timing constraints, i.e., $t_{s} + \Delta t< t_{mission}$.
Collision occurs only if the global minimum is found to be non-positive.

\begin{figure}
	\begin{center}
		\includegraphics[width=0.48\textwidth]{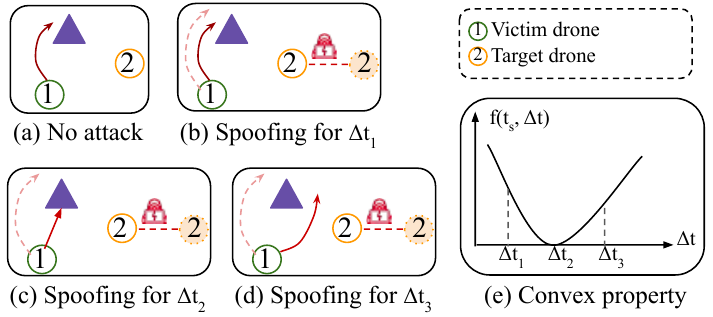}
	\end{center}
	\caption{Convex property of the objective function in \sysOld.}
	\label{fig:convex} 
\end{figure}

We make the observation that the objective function $f$ {\em is convex} (Fig.~\ref{fig:convex}-(e)).
This is because performing spoofing for either too short (e.g., $\Delta t_{1}$) or too long (e.g., $\Delta t_{3}$) a time will cause the victim drone to avoid the obstacle from either side.

For example, the victim drone avoids the obstacle from the left side in the absence of the attack (Fig.~\ref{fig:convex}-(a)). A too short spoofing time $\Delta t_{1}$ would  make it miss the obstacle from the left (Fig.~\ref{fig:convex}-(b)), 
When the time increases to $\Delta t_{2}$ (Fig.~\ref{fig:convex}-(c)), it causes a collision.
However, if the spoofing time is too long and increases to $\Delta t_{3}$, 
the victim drone avoids the obstacle from the right (Fig.~\ref{fig:convex}-(d)), 
thus increasing the value of $f({t_s}, \Delta t)$ (Fig.~\ref{fig:convex}-(e)). Thus, we need to find a spoofing time that is neither too short nor too long, to cause the collision. 

Since the objective function is convex, we use gradient-guided search to find an optimal solution. \newtcps{The algorithm's details can be found in Section IV-C in the original SwarmFuzz paper~\cite{swarmfuzz}.}

\section{Evaluation of \sysOld}

First, we present our experimental setup, and then the results for effectiveness of \sysOld in various swarm configurations. 
Finally, we perform an ablation study of \sysOld's heuristics.  

\subsection{Experimental Setup}
\label{sec:experiment}
\textbf{Simulator.} 
We use SwarmLab~\cite{swarmlab}, a popular swarm simulator, that implements drone dynamics accurately~\cite{swarmlab}, and also implements state-of-the-art control algorithms. 
\new{To evaluate \sysOld, we choose two swarm control algorithms implemented by the simulator. They are Vicsek algorithm~\cite{Viscek} and Olfati-Saber algorithm~\cite{OlfatiSaber}.
Vicsek algorithm was published in 2018 and Olfati-Saber algorithm was published in 2002.
The Vicsek algorithm has been cited $365$\cite{cite-vicsek} times  and the Olfati-Saber algorithm has been cited $856$\cite{cite-olfati} times (as of August 2023).
The Vicsek algorithm focused on the alignment of individual swarm members' velocities to create collective group behavior. It has been validated on real hardware with a $30$ drone swarm, and performs well in terms of maintaining the formation as well as obstacle avoidance \cite{Viscek}. 
In contrast, the Olfati-Saber algorithm leverages potential field methods \cite{pfm} to 
model the interactions between agents and the environment as attractive and repulsive forces.
These virtual forces are then used to generate velocities to guide drones away from the obstacle and toward unexplored regions.}
Both algorithms perform collision avoidance based solely on the GPS sensor readings.
For brevity, we henceforth refer to the Vicsek algorithm as "A1" and the Olfati-Saber algorithm as "A2". 

Each drone in SwarmLab is a quadcopter weighted at $0.296$kg (by default), and using a PID (Proportional-Integral-Derivative) flight controller.

\textbf{GPS Spoofing.}
As we did not have access to GPS transmitter hardware, we simulate GPS spoofing attacks through software code modification - this is similar to what a lot of prior work has done~\cite{pidpiper, stealthy-attacks}. 
We launch GPS spoofing in software by manipulating the GPS reading to $GPS+d$ at the GPS sampling rate (100 Hz in SwarmLab by default), where $d$ is the spoofing deviation. We consider $d$ to be $5m$ and $10m$ respectively. 
Most defense techniques \cite{pidpiper, semperfi} ignore spoofing distances of less than $10m$ as such small deviations are indistinguishable from standard GPS offset \cite{gps-offset}, in order to avoid false-positives. 
Therefore, we inject only small amounts of noise (i.e., $5m/10m$) during GPS spoofing as this is hard to detect by current defense techniques, and thus is stealthy.
Even with this small GPS spoofing distance, we show that the attacker can achieve her objective with a high success rate. 

\textbf{Mission Details.}
\new{Since the objectives of the two algorithms are different, we consider two types of missions, one for each algorithm. 
By doing so, we can assess the resilience of the swarm control algorithms 
and highlight the limitations of each algorithm in their intended applications.}

\new{Vicsek algorithm is designed to navigate the drone swarms to a destination. Therefore, it is suitable for scenarios where the primary goal is to reach a specific destination efficiently. The Vicsek algorithm excels in tasks that require coordinated movement, such as formation flying, rather than collision avoidance. Therefore, we consider the whole phase of a mission that aims to reach a pre-defined destination ($233.5m$ away), 
while avoiding a single on-path obstacle. The obstacle is placed at roughly the halfway mark of the mission so that the drone swarm has enough time to react for collision avoidance. Because the obstacle is known to the mission, the drone swarm can avoid colliding with the obstacle under normal operation.}

\new{On the other hand, Olfati-Saber algorithm is designed to explore the surrounding environment while avoiding obstacles. This algorithm prioritizes exploration and obstacle avoidance, making it suitable for complex missions with multiple obstacles. Therefore, we consider the entire mission phase, which includes $14$ obstacles spaced $50$ meters apart within a confined $300m*300m$ area. Each instance of the two types of missions in SwarmLab takes around $120s$ to finish on average.}

To reduce the bias, the initial location of the drone swarm is randomly generated within a range of $0-50m$ relative to the mission starting point.
We limit the range to $50m$ as the drone swarm is sparse even with a large size (e.g., 15 drones) and is unlikely to have collisions under attack.

\textbf{Success Metric.}
In the absence of attacks, we find that no collision occurs in any mission.
We consider an SPV to be successfully found, \new{if either (1) the victim drone is not making progress towards the destination (i.e., freezes), or (2) collision occurs involving the victim drone(s). }
Note that we do not consider collisions caused directly by the target drone (i.e., the target drone collides with the victim or the obstacle).

\subsection{Effectiveness of \sysOld across Swarm Configurations}
\label{sec: effectiveness}
We apply \sysOld to six drone swarm configurations, with swarm sizes of 5, 10, 15 drones, and GPS spoofing distances of $5m$ and $10m$. 
We perform 100 missions for each configuration, to obtain a representative sample. 
We also manually validate each vulnerability found by SwarmFuzz via simulation, and find that all of them are True Positives (TP).
Then we measure in how many missions \sysOld found \atkname - this is the success rate of \sysOld for that configuration. 
An alternate way to measure success rate is the ratio of the number of  \atkname found to the maximum number of   \atkname for a given configuration.
However, we cannot easily obtain the maximum possible number of \atkname as this requires exhaustive sampling of the input space, which is prohibitively expensive. 

\new{Unfortunately, \sysOld fails to find any SPVs in A2. This is because most seeds selected cannot lead to attack scenarios. Therefore, we only report the results for A1. In Section~\ref{sec:methdology_new}, we propose a variant of \sysOld, \sysNew{} that works for both algorithms.}

\begin{table}[!ht]
	\caption{Success rates of \sysOld in finding \atkname in A1.}
	\label{table: success}
	\centering
	\begin{tabular}{|l|l|l|l|}
		\hline
		Swarm size & 5 drones & 10 drones & 15 drones \\ \hline
		5m spoofing & 21\% & 36\% & 54\% \\ \hline
		10m spoofing & 49\% & 59\% & 74\% \\ \hline
	\end{tabular}
\end{table}

\new{Overall, \sysOld finds that the \atkname all lead to collisions between the victim drone and the on-path obstacle.} Table~\ref{table: success} shows the results. We observe that success rates of \sysOld in A1 vary from 21\% to 74\% across different configurations (average $48.8\%$ 
). \emph{This shows that \sysOld is highly  effective in finding \atkname for different swarm configurations. }

Further, \sysOld has a higher success rate for missions when the (1) GPS spoofing distance is higher, as it allows the attacker to disrupt the swarm more, and 
 (2) when the swarm size is higher, as larger swarms are denser, making obstacle avoidance more difficult. \newtcps{The detailed analysis can be found in Section V-B in the original SwarmFuzz paper~\cite{swarmfuzz}.}

\begin{figure}[ht] 
	\begin{subfigure}[b]{0.5\linewidth}
		\centering
		\includegraphics[width=0.95\linewidth]{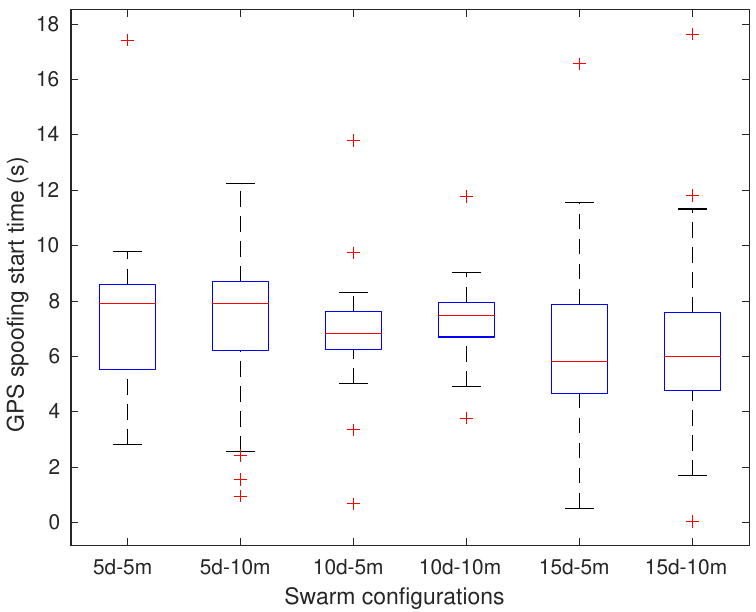} 
		\caption{GPS spoofing starting time. } 
		\label{fig: RQ1-GPS-start} 
	\end{subfigure}
	\begin{subfigure}[b]{0.5\linewidth}
		\centering
		\includegraphics[width=0.95\linewidth]{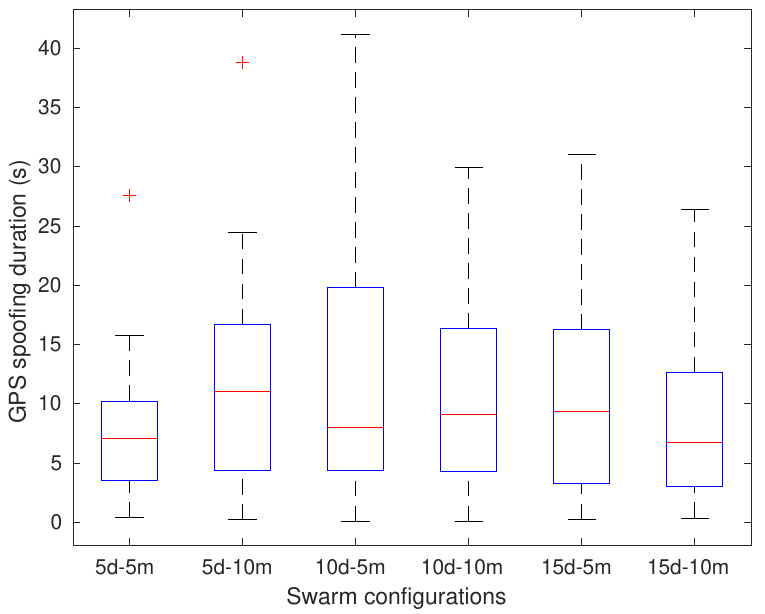} 
		\caption{GPS spoofing duration. } 
		\label{fig: RQ1-GPS-duration} 
	\end{subfigure} 
	\caption{GPS spoofing parameters found by \sysOld across swarm configurations. In the figure, 
	"5d-5m" means 5-drone swarms under 5m-spoofing, and so on.}
	\label{fig: GPS params} 
\end{figure}

Fig.~\ref{fig: GPS params} shows the GPS spoofing parameters (i.e., the starting time and the duration) to trigger the \atkname under different swarm configurations.
These are found by SwarmFuzz during the gradient-based optimization process. 
We find that the average GPS spoofing starting time across different configurations is $6.91s$, 
and the average GPS spoofing duration is $10.33s$.

We also report the average number of search iterations across different swarm configurations in Table~\ref{table: V_RQ1_fuzzing_time}.  
Each iteration takes 120s (the mission time), and hence this is correlated with the runtime.
We find that the average runtime increases with the size of the drone swarm, since the interactions among drones become more complex.
For a given swarm size, the runtime overheads  under different GPS spoofing deviations exhibit no significant differences across configurations.

\begin{table}[]
	\caption{Average number of search iterations for \sysOld to find \atkname across swarm configurations in A1.}
	\centering
	\label{table: V_RQ1_fuzzing_time}
	\begin{tabular}{|l|l|l|l|}
		\hline
		& 5-drone & 10-drone & 15-drone \\ \hline
		5m-spoofing & 6.33 & 9.3 & 12.65  \\ \hline
		10m-spoofing & 6.93 & 9.91 & 13.47  \\ \hline
	\end{tabular}
\end{table}

\subsection{Ablation Study: Effectiveness of \sysOld's Heuristics} 
\label{sec:ablation1}
To the best of our knowledge, there is no existing fuzzing tool to find SPVs in drone swarms, and thus there is no prior work we can use for comparison with \sysOld. 
Instead, we perform an ablation study and compare \sysOld with different variants of itself, and with random fuzzing. 

We used two heuristics to improve the efficiency of \sysOld in Section~\ref{sec:methdology}, 
(1) using the \graphname to prioritize the influential drone pairs selected for fuzzing;
(2) using gradient-guided optimization to find spoofing parameters for causing collisions.
To evaluate the effect of these heuristics, we compare \sysOld with three other fuzzers as follows. 
(1) R\_Fuzz does not implement either heuristic,  and instead randomly chooses the drone pairs as well as the spoofing parameters. 
(2) G\_Fuzz only implements the gradient-guided optimization to search for spoofing parameters but not the \graphname, instead choosing these pairs randomly.
(3) S\_Fuzz only implements the \graphname to choose the drones pairs but not the gradient-guided optimization, instead choosing the spoofing parameters randomly.
Therefore, we evaluate the efficacy of the \graphname by comparing \sysOld with G\_Fuzz, 
and the efficacy of gradient-guided optimization by comparing it with S\_Fuzz.

We measure two metrics for each fuzzer, the success rate and the runtime overhead.
For runtime overhead, we report the average number of search iterations taken by the fuzzer - each iteration takes about 2 minutes (time taken by a mission). 

We also cap the number of search iterations for each seed to $20$ in all the fuzzers
based on our empirical observations.
We experimentally observe that the number of  vulnerabilities found saturates at $20$ search iterations, which corresponds to  $40$ min.
We evaluate the fuzzers with the 5-drone swarm with a $10m$ GPS spoofing distance. Table~\ref{table: V_RQ2_comp} shows the results. 

\begin{table}[]
	\caption{Comparison of fuzzers in 5 drones, 10m spoofing.}
	\vspace{-2mm}
	\centering
	\label{table: V_RQ2_comp}
	\begin{tabular}{|l|l|l|l|l|}
		\hline
		& \sysOld & R\_Fuzz & G\_Fuzz & S\_Fuzz\\ \hline
		Success rate & 49\% & 8\% & 5\% & 12\%  \\ \hline
		Avg. iterations & 6.93 & 19.52 & 6.75 & 19.85\\ \hline
	\end{tabular}
\end{table}

We find that the success rate of \sysOld is $49\%$, whereas the success rate of R\_Fuzz is just $8\%$, and that of the G\_Fuzz just $5\%$. 
Further, the  runtime overhead of \sysOld and  G\_Fuzz is low (about $7$ iterations), unlike those of R\_Fuzz and S\_Fuzz (about $20$ iterations, i.e., maximum number of search iterations). Compared to G\_Fuzz, the success rate of \sysOld is about $10$x higher, while compared to S\_Fuzz, its runtime overhead is $3$x lower. 
Thus, the \graphname boosts the success rate of \sysOld by almost $10$x, while the gradient-guided optimization reduces its runtime overhead  by about $3$x. Compared to the random fuzzer (R\_Fuzz), \sysOld has about $6$x higher success rate and 3x smaller runtime overhead.

\section{Methodology of \sysNew{}}
\label{sec:methdology_new}
\new{As mentioned in Section~\ref{sec: effectiveness}, \sysOld{} cannot find any \atkname in A2. 
We find this is due to two reasons, (1) lack of generality of graph-based seed scheduling and (2) overshooting of the target by the gradient descent algorithm.}

\new{For limitation 1, \sysOld{} calculated malicious influence from the constructed graph and used this influence to select target-victim drone pairs. However, if the graph fails to accurately model the swarm's topology, the computed influence may be incorrect. As a result, the selected drone pairs may not effectively trigger \atkname, leading to ineffective seed scheduling.}


\new{For limitation 2, the efficiency of the gradient descent algorithm heavily depends on the learning rate. A high learning rate enables faster convergence but increases the risk of overshooting, while a low learning rate ensures more precise convergence but slows down the process.  Therefore, the learning rate must be carefully tuned for each objective function.}

\begin{figure}[!ht]
	\begin{center}
		\includegraphics[width=0.5\textwidth]{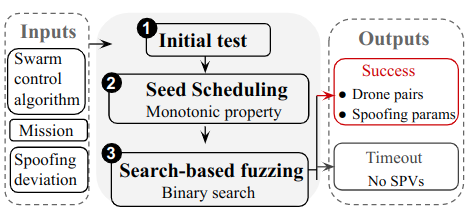}
	\end{center}
	\caption{\new{Overview of \sysNew{}}}
	\label{fig: overview-Bin} 
\end{figure}

\new{To overcome the above limitations, we propose \sysNew{}, a new framework based on \sysOld{}. 
Fig.~\ref{fig: overview-Bin}  presents the overview of \sysNew{}.
We modify the step 2 and step 3 in \sysOld{}.
In step 2, seed scheduling, we apply the property of the monotonic function, to identify the 
malicious influence of the drone pairs. This overcomes limitation 1 as it is topology-agnostic.
In step 3, we apply binary search, instead of the gradient-based method, to find the SPV cases in the search-based fuzzing. This overcomes limitation 2 as binary search is parameter-agnostic.}

\new{\sysNew takes as input the drone swarm and a set of GPS spoofing parameter options. These parameters are used to get the seed pool from Algorithm~\ref{alg:seed} (Line 3). \sysNew goes through each seed, calls \textproc{BinSearchDuration} (Algorithm~\ref{alg:binary}) to obtain the successful GPS spoofing duration $\hat{T_i}$ and $VDO$ (Line 5). If an SPV is found,  we update the involved GPS spoofing parameters.}

\subsection{Seed scheduling}
\label{sec: new scheduling}

\new{The goal of seed scheduling is to prioritize the most influential drone pairs during fuzzing. Instead of quantifying influence as in Section~\ref{IV.B}, we qualitatively assess the malicious influence of a drone pair. This influence is inferred by comparing two executions: one with the maximum attack applied and one without any attack. The behavioral differences observed in the victim drone between these executions indicate the range of the attack's impact.}

\new{We observe that if the maximum attack fails to cause a collision, then a weaker attack will also be ineffective. Based on this insight, we transform the convex VDO function from Section~\ref{sec: convex} into a monotonic function. Using the property of monotonic functions, where a zero exists if the function’s values at the maximum and minimum points have opposite signs, we determine whether an attack is possible.}

\begin{figure}[!ht]
	\begin{center}
		\includegraphics[width=0.5\textwidth]{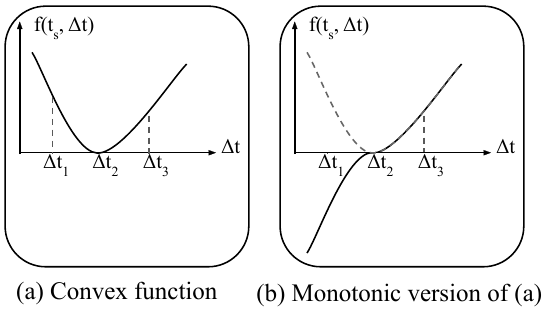}
	\end{center}
	\caption{\new{VDO function in the convex and the monotonic forms.}}
	\label{fig:mono} 
\end{figure}

\new{To transform the VDO function in Section~\ref{sec: convex} into a monotonic one, we flip part of the function by adding a sign (shown in Fig.~\ref{fig:mono}-(b)). The sign is decided based on the direction that the victim drone passes the on-path object. Specifically, if the victim drone passes the on-path object from the right, the sign is positive, otherwise, the sign is negative. In this monotonic function, a collision (e.g., $f({\Delta}t_{2})=0$) is only possible, if the signs of the victim drone passing the on-path object under no attack are different from that under the maximum attack (e.g., $f({\Delta}t_{1})*f({\Delta}t_{3})<0$). 
Therefore, if these opposite signs are found, we infer the target drone has a malicious influence, and add the target-victim drone pair into the seed pool.}

\new{Algorithm ~\ref{alg:seed} describes the \textproc{SeedScheduling} function. The seed scheduling takes as input a set of GPS spoofing parameter options. The parameters include the GPS spoofing distance, direction, start time, and duration. To exercise the victim drone's behavior under different maximum amounts of GPS spoofing, the parameter options include performing GPS spoofing during the whole mission to deviate the victim drone (1) to its right and (2) to its left. \textproc{SeedScheduling} goes through every possible target-victim drone pair (Line 3), gets the victim drone's direction passing the on-path object without attack (Line 4), and such direction under different maximum GPS spoofing attack scenarios (Line 6-7). If the victim drone changes its direction to pass the on-path object under the GPS spoofing attack (Line 8), then the target drone has the malicious influence on the victim drone. Finally, we add such drone pairs and the involved GPS spoofing parameters to the seed pool.}

\begin{algorithm}
\caption{\new{The \textproc{SeedScheduling} function.}}\label{alg:seed}
\textbf{Input:} GPS spoofing parameter options $\Theta=\{\theta_{i}\}_{i}$ \\
\textbf{Output:} seed pool $S$.
\begin{algorithmic}[1]
\Function{SeedScheduling}{$\Theta$}
\State $S \gets \emptyset$
\For{every possible drone pair $(x_{t}, x_{v})$}
\State $V_{before} \in \{\pm{1}\} \gets \textproc{GetRelativePos}(x_{t}, x_{v})$
\For {$\theta_{i} \in \Theta$}
\State $\hat{x}_{t} \gets \textproc{Spoof}(x_{t}, \theta_{i})$
\State $V_{after} \in \{\pm{1}\} \gets \textproc{GetRelativePos}(\hat{x}_{t}, x_{v})$
\If{$V_{before} = - V_{after}$}
\State $S \gets S \cup \{(x_{t}, x_{v}, \theta_{i})\}$
\EndIf
\EndFor
\EndFor
\\
\Return $S$
\EndFunction
\end{algorithmic}
\end{algorithm}

\subsection{Search-based fuzzing}
\new{The goal of search-based fuzzing is to find attack parameters that cause collisions in the swarm. We model this as an optimization problem, i.e., to minimize the distance between the victim drone and the on-path object. 
Since the objective function has been transformed into its monotonic form (Fig.~\ref{fig:mono}), we use binary search to efficiently find such attack parameters.
The attack parameters consist of GPS spoofing start time and duration.
We choose the spoofing start time as $0s$, as this does not require the attacker to track the swarm during the mission and is hence easier. 
Therefore, the goal is to find the GPS spoofing duration that causes the collision.}

\new{Algorithm ~\ref{alg:binary} describes the \textproc{BinSearchDuration} function.
The goal of the \textproc{BinSearchDuration} function is to
find the GPS spoofing duration $\hat{T}$ that will trigger \atkname and the corresponding $VDO$, given the seed $(x_{t}, x_{v}, d_i, D_i, t_i, T_i)$.}

\new{\textproc{BinSearchDuration} starts by initializing the end searching points ($T_{min}$ and $T_{max}$) and the VDO (Line 2). 
Then it calculates the victim drone's direction of passing the on-path object under no attack ($V_{min}$ in Line 3) and under maximum attack ($V_{max}$ in Line 4-5), respectively.
As long as the condition for the binary search holds (Line 6), \textproc{BinSearchDuration} calculates the mid-point function value (Line 7-9) and the corresponding VDO (Line 10). An SPV is found if the $VDO$ is smaller than the radius of the drone size (Line 11-13). 
If not, \textproc{BinSearchDuration} updates the end searching point ($T_{min}/T_{max}$) and the victim drone's direction ($V_{min}/V_{max}$) accordingly (Line 15-19).}


\begin{algorithm}
\caption{\new{The \textproc{BinSearchDuration} function.}}\label{alg:binary}
\textbf{Input:}  target drone $x_{t}$, 
                 victim drone $x_{v}$, 
                 GPS spoofing deviation $d_i$, 
                 GPS spoofing direction $D_i$, 
                 GPS spoofing start time $t_i$, 
                 GPS spoofing duration $T_i$,
                 Radius of the drone size $\epsilon$.\\
\textbf{Output:} Successful spoofing duration $\hat{T}$, Victim drone's distance to the on-path object $VDO$. 
\begin{algorithmic}[1]
\Function{BinSearch}{$x_{t}, x_{v}, d_i, D_i, t_i, T_i$}
\State $T_{min} \gets 0, T_{max} \gets \Tilde{T
_i}, VDO \gets +\infty$
\State $V_{min}\in \{\pm{1}\} \gets \textproc{GetDirection}(x_{t}, x_{v})$ 
\State $\hat{x}_{t} \gets \textproc{Spoof}(x_{t}, (d, D, t, \Tilde{T}))$, 
\State $V_{max} \in \{\pm{1}\} \gets \textproc{GetDirection}(\hat{x}_{t}, x_{v})$
\While{$T_{max}>T_{min}$}
\State $T_{mid} = (T_{min}+T_{max})/2$
\State $\hat{x}_{t} \gets \textproc{Spoof}(x_{t}, (d, D, t, T_{mid}))$
\State $V_{mid} \in \{\pm{1}\} \gets \textproc{GetDirection}(\hat{x}_t, x_{v})$
\State $VDO \gets \textproc{GetVDO}(\hat{x}_{t}, x_{v})$
\If{$|VDO| \leq \epsilon$}
\State $\hat{T} \gets T_{mid}$
\State break
\Else
\If{$V_{mid}=V_{min}$}
\State $T_{min} \gets T_{mid}, \ V_{min} \gets -V_{mid}$
\Else
\State $T_{max} \gets T_{mid}, \ V_{max} \gets -V_{mid}$
\EndIf
\EndIf
\EndWhile\\
\Return $\hat{T}$, $VDO$
\EndFunction
\end{algorithmic}
\end{algorithm}

\section{Evaluation of \sysNew}

\new{The experimental setup is the same as that in Section~\ref{sec:experiment}. We first present the results for the effectiveness of \sysNew in various swarm configurations. 
Then, we perform an ablation study of \sysNew's heuristics.}

\subsection{Effectiveness of \sysNew across Swarm Configurations}
\new{We apply \sysNew to six drone swarm configurations, consisting of swarm sizes of 5, 10, 15 drones, and GPS spoofing distances of $5m$ and $10m$. 
We perform 100 missions for each configuration, to obtain a representative sample.}


\begin{table}[ht]
\centering
\begin{tabular}{|c|c|c|c|c|}
\hline
\multirow{2}{*}{Swarm size} & \multicolumn{4}{c|}{Success Rate (\%)} \\
\cline{2-5}
 & 5 drones & 10 drones & 15 drones &  Average\\
\hline
\multirow{2}{*}{5m spoofing} & 28 (A1) & 42 (A1) & 83 (A1) & 51 (A1)\\
 & 61 (A2) & 87 (A2) & 97 (A2) & 81.7 (A2)\\
\hline
\multirow{2}{*}{10m spoofing} & 49 (A1) & 65 (A1) & 94 (A1) & 69 (A1) \\
 & 75 (A2) & 99 (A2) & 100 (A2) & 91.3 (A2) \\
\hline
\end{tabular}
\caption{\new{Success rates of \sysNew in finding SPVs in A1 and A2.}}
\label{tab:success-1}
\end{table}

\new{\textbf{Success rates. }
Table~\ref{tab:success-1} shows the results. 
We observe that success rates of \sysNew vary from 28\% to 100\% across different configurations and swarm control algorithms. \sysNew also has a higher success rate in finding \atkname in A2 (Avg. $86.5\%$) than in A1 (Avg. $60.2\%$).
This is because the collision scenarios in A2 are more likely to occur compared to those in A1. In A1, collisions only occur between the drone and the obstacle, whereas, in A2, collisions can also occur between the drones within the swarm. For example, in a 5-drone swarm with one on-path obstacle, the number of combinations of collisions objects in A1 are $5*1 = 5$. However, in A2, this number increases into ${5 \choose 2} = 10$. }

\new{
We also compare the success rates of \sysNew and \sysOld in A1. We exclude A2 from this comparison as \sysOld did not find \atkname in A2. 
Table~\ref{tab:success-A1} shows the results.
"G" denotes the results for \sysOld, while "B" denotes the results for \sysNew. Overall, we find the success rate of \sysNew is higher than that of \sysOld.
\emph{These results show that \sysNew is highly effective in finding \atkname for different swarm configurations. }}

\begin{table}[ht]
\centering
\begin{tabular}{|c|c|c|c|c|}
\hline
\multirow{2}{*}{Swarm size} & \multicolumn{4}{c|}{Success Rate (\%)} \\
\cline{2-5}
 & 5 drones & 10 drones & 15 drones &  Average\\
\hline
\multirow{2}{*}{5m spoofing} & 21 (G) & 36 (G) & 54 (G) & 37 (G)\\
 & 28 (B) & 42 (B) & 83 (B) & 51 (B)\\
\hline
\multirow{2}{*}{10m spoofing} & 49 (G) & 59 (G) & 74 (G) & 60 (G) \\
 & 49 (B) & 65 (B) & 94 (B) & 69 (B) \\
\hline
\end{tabular}
\caption{\new{Success rates of \sysNew(B) vs. \sysOld (G) in finding SPVs in A1.}}
\label{tab:success-A1}
\end{table}

\new{Further, similar to Section~\ref{sec:experiment}, \sysNew has a higher success rate for missions when the (1) GPS spoofing distance is higher, as it allows the attacker to disrupt the swarm more, and  
(2) when the swarm size is higher, as larger swarms are denser, making collision avoidance between drones more difficult.}

\begin{figure}[ht] 
	\begin{subfigure}[b]{0.5\linewidth}
		\centering
		\includegraphics[width=0.8\linewidth]{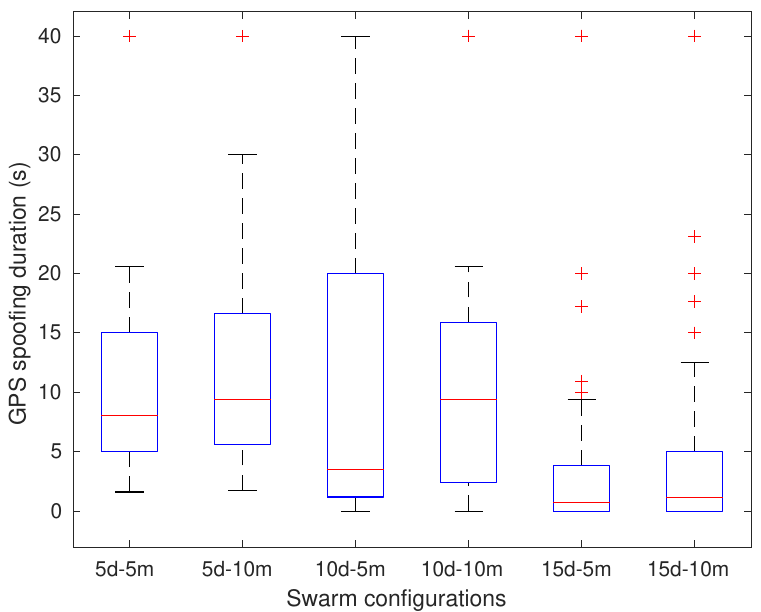} 
		\caption{A1. } 
		\label{fig: algo1-GPS-duration} 
	\end{subfigure}
	\begin{subfigure}[b]{0.5\linewidth}
		\centering
		\includegraphics[width=0.8\linewidth]{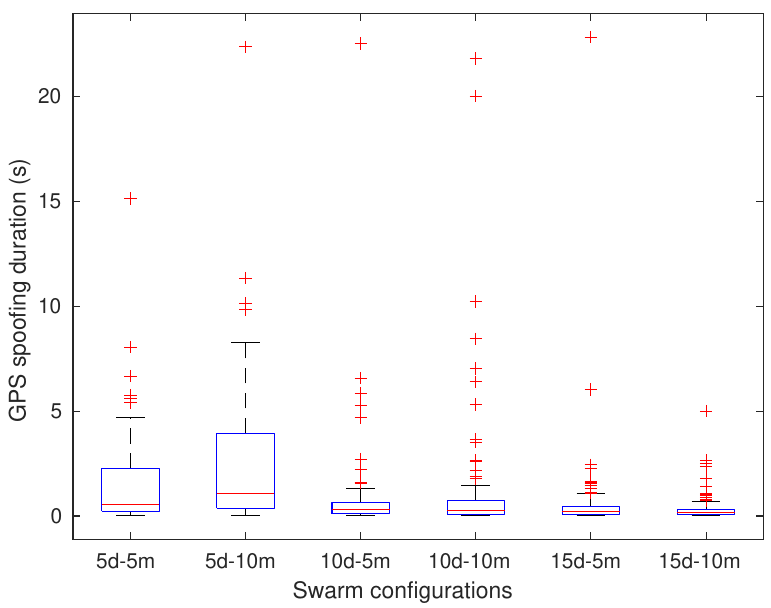} 
		\caption{A2. } 
		\label{fig: algo2-GPS-duration} 
	\end{subfigure} 
	\caption{GPS spoofing duration found by \sysNew across swarm configurations in algorithms A1 and A2. In the figure, 
		"5d-5m" means 5-drone swarms under 5m-spoofing. }
	\label{fig:algo2-GPS_params}  
\end{figure}

\new{\textbf{Attack parameters. }Fig.~\ref{fig:algo2-GPS_params} shows the GPS spoofing duration to trigger the \atkname under different swarm configurations. These are found by \sysNew during the binary search process. As for the GPS spoofing start time, we choose $0s$ by default. This saves attacker efforts to track the swarm during the whole mission.
We find that the average GPS spoofing duration across different configurations for A2 (Avg. $1.15s$) is much lower than that for A1 (Avg. $8s$). Thus, A2 is more subject to small location disruptions due to GPS spoofing. 
We also observe that for both algorithms,  with larger swarm sizes, the GPS spoofing duration is shorter, except in the 10d-5m setting.
This is because larger swarms are denser, making collision avoidance between drones already more difficult.}

\begin{figure}[ht] 
	\begin{subfigure}[b]{0.5\linewidth}
		\centering
		\includegraphics[width=0.8\linewidth]{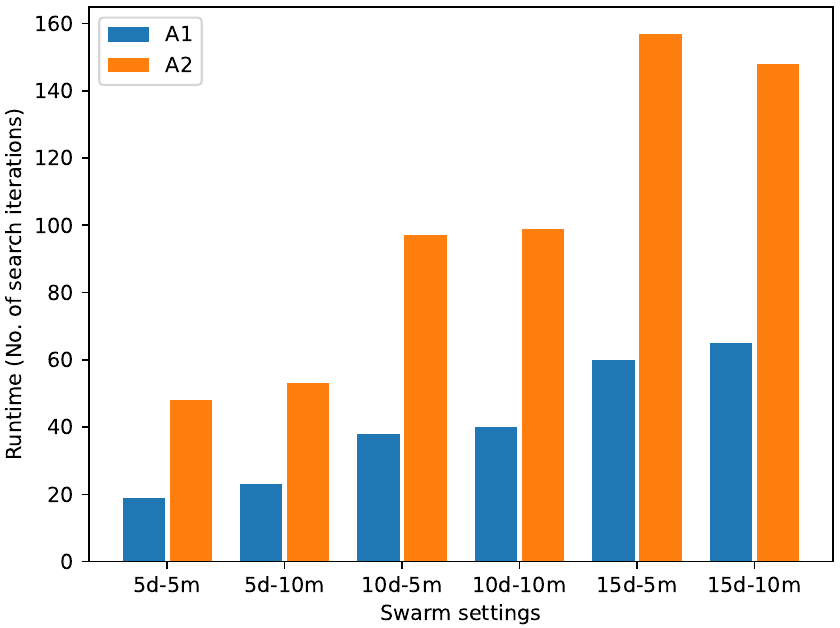} 
		\caption{\sysNew's overhead in A1 and A2. } 
		\label{fig:overhead-1} 
	\end{subfigure}
	\begin{subfigure}[b]{0.5\linewidth}
		\centering
		\includegraphics[width=0.8\linewidth]{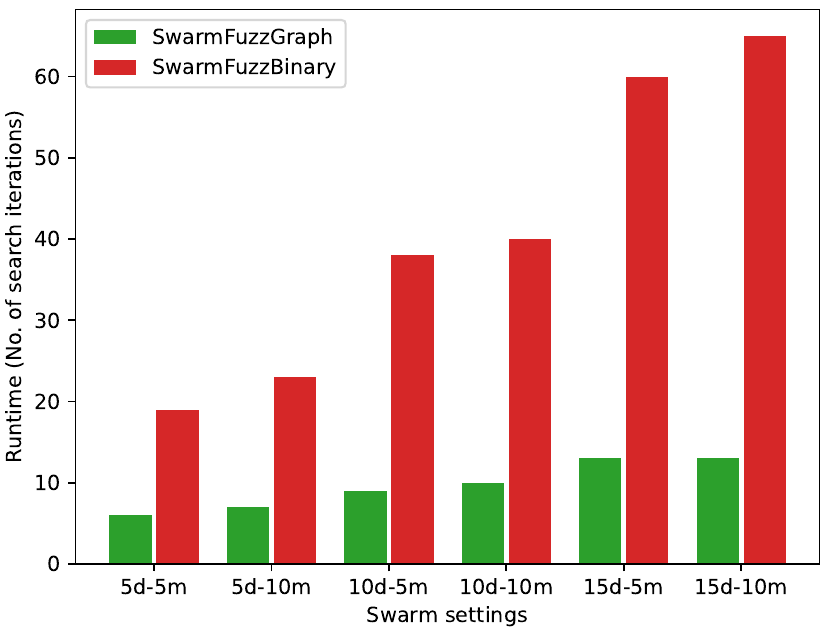} 
		\caption{\sysNew vs. \sysOld in A1. } 
		\label{fig:overhead-2} 
	\end{subfigure} 
	\caption{\new{Average number of search iterations taken to find \atkname across swarm configurations.}}
	\label{fig:overhead-all}  
\end{figure}

\new{\textbf{Runtime. }We also report the average number of search iterations across different swarm configurations in Figure~\ref{fig:overhead-1}.  
Within the swarm control algorithm, we find that the average runtime increases with the size of the drone swarm, since the interactions among drones become more complex.
For a given swarm size, the runtime under different GPS spoofing deviations exhibits no significant differences across configurations. 
When comparing the swarm control algorithms, we find that the runtime of \sysNew on A2 is $2.5x$ higher than that on A1.
This is because, for each mission, \sysNew generates more seeds in A2 than in A1 
and hence, the runtime of \sysNew on A2 is higher.
We further find that the number of seeds generated is related to the type of collision caused by \atkname. 
In A1, collisions only occur between the victim drone and the obstacle,
whereas, in A2, collisions occur between two victim drones. } 


\new{We also compare the average number of search iterations taken by \sysNew 
with that taken by \sysOld 
to find the \atkname. Figure~\ref{fig:overhead-2} shows the results. We find that on average, the runtime of \sysNew is 4.2x higher than that of \sysOld. This high runtime is mainly due to the difference in the seed scheduling in each framework. 
\sysOld only needs to run one simulation to construct the \graphname, and then performs the seed scheduling in a single shot. 
However, \sysNew needs to iterate on every possible combination of the target-victim drone pairs to choose the seeds.} 



\begin{figure}[ht] 
	\begin{subfigure}[b]{0.5\linewidth}
		\centering
		\includegraphics[width=0.8\linewidth]{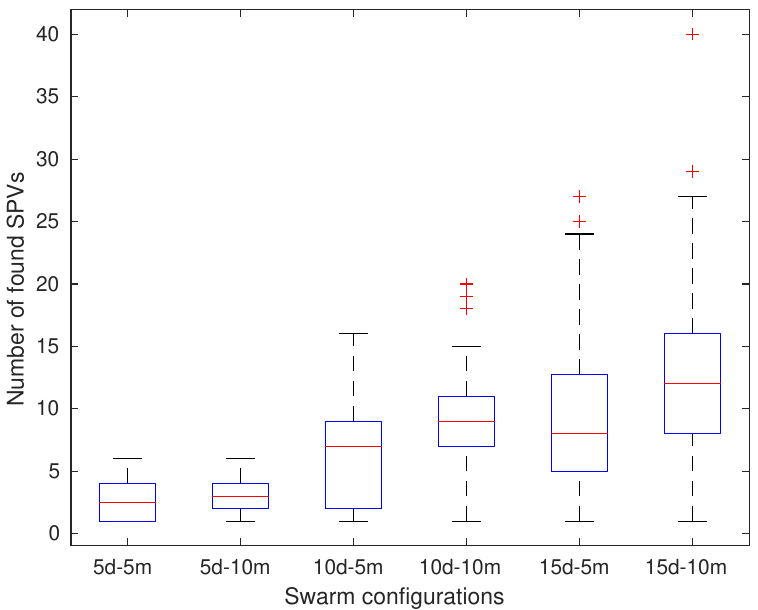} 
		\caption{A1. } 
		\label{fig: algo1-nbSPVs} 
	\end{subfigure}
	\begin{subfigure}[b]{0.5\linewidth}
		\centering
		\includegraphics[width=0.8\linewidth]{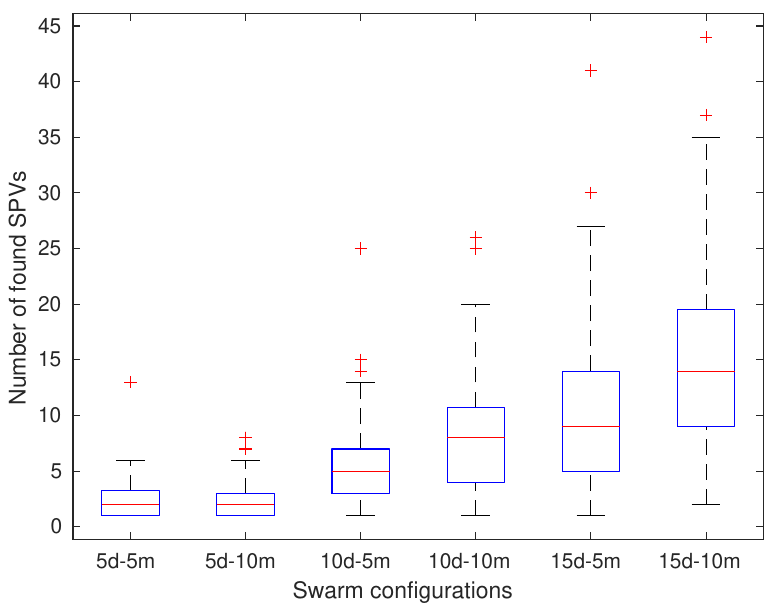} 
		\caption{A2. } 
		\label{fig: algo2-nbSPVs} 
	\end{subfigure} 
	\caption{\new{Boxplot for the number of found SPVs per mission in algorithm A1 and A2. In the figure, 
		"5d-5m" means 5-drone swarms under 5m-spoofing, and so on.}}
	\label{fig:nb-SPVs}  
\end{figure}

\new{\textbf{Number of SPVs found. } We observe that the number of found \atkname differs across algorithms. On average, each A1 mission has $5$ SPVs, while each A2 mission has $7$ SPVs. Fig. ~\ref{fig:nb-SPVs} shows the number of SPVs found in each mission setting. We observe a similar trend in the number of SPVs found as the success rate, i.e., the number of found \atkname increases with larger swarm sizes and higher spoofing distances.}

\subsection{Ablation study}


\new{As we did with \sysOld, to evaluate the effect of heuristics, we perform an ablation study and compare \sysNew with three other fuzzers as shown in table~\ref{table: fuzzers}}


\begin{table}[]
	\caption{\new{Fuzzers for comparision.}}
	\label{table: fuzzers}
	\begin{tabular}{|l|l|l|}
		\hline
		& Seed scheduling & Binary search \\ \hline
	       \sysNew & \cmark  & \cmark \\ \hline
		R\_Fuzz & \xmark & \xmark \\ \hline
		B\_Fuzz & \xmark & \cmark \\ \hline
		S\_Fuzz & \cmark & \xmark \\ \hline
	\end{tabular}
\end{table}

\begin{figure}[ht] 
    \centering
    \begin{subfigure}[b]{0.45\linewidth}
        \centering
        \includegraphics[width=0.7\linewidth]{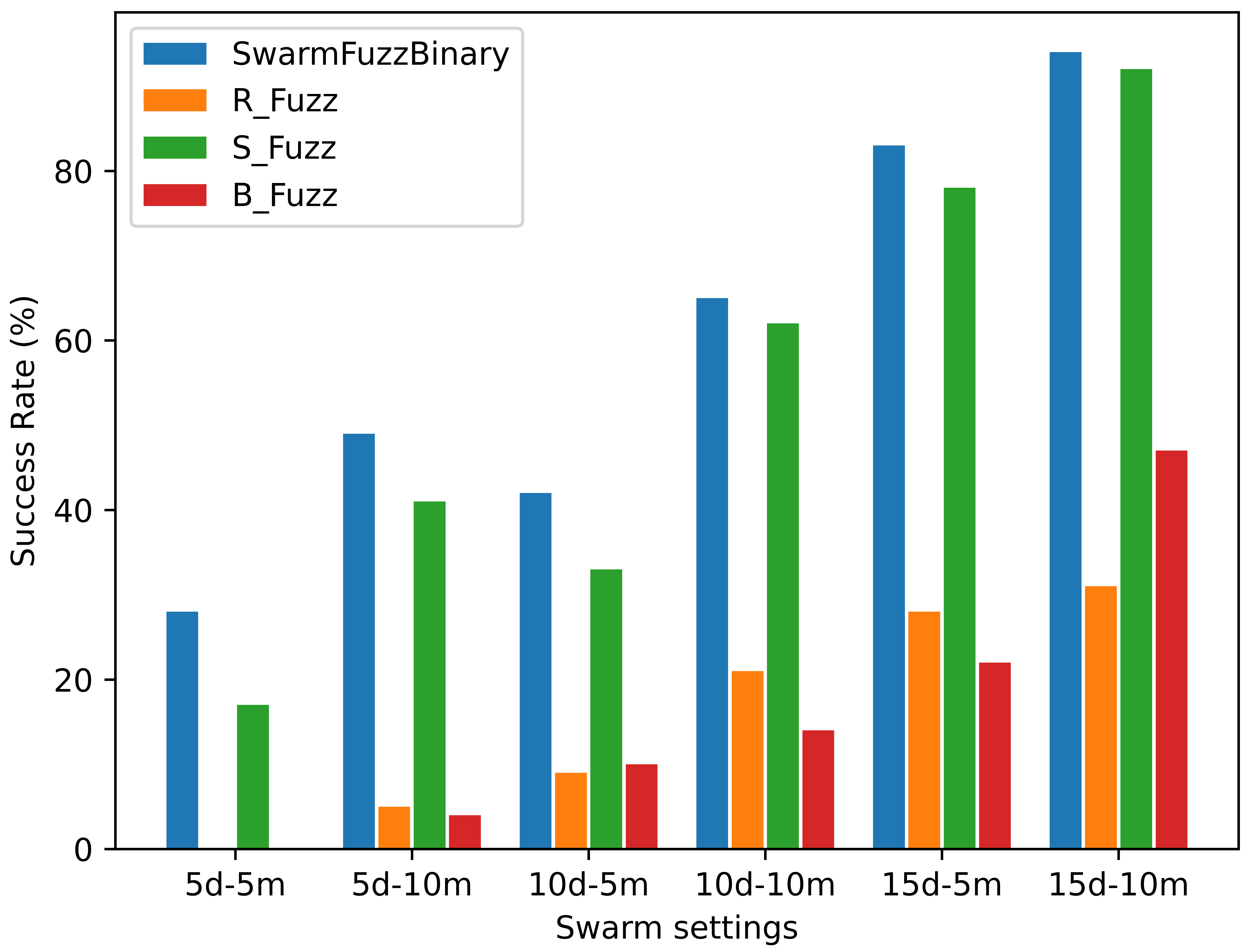} 
        \caption{Success rate (A1)}
        \label{fig:algo2-ablation-suc-A1} 
    \end{subfigure}
    \hfill
    \begin{subfigure}[b]{0.45\linewidth}
        \centering
        \includegraphics[width=0.7\linewidth]{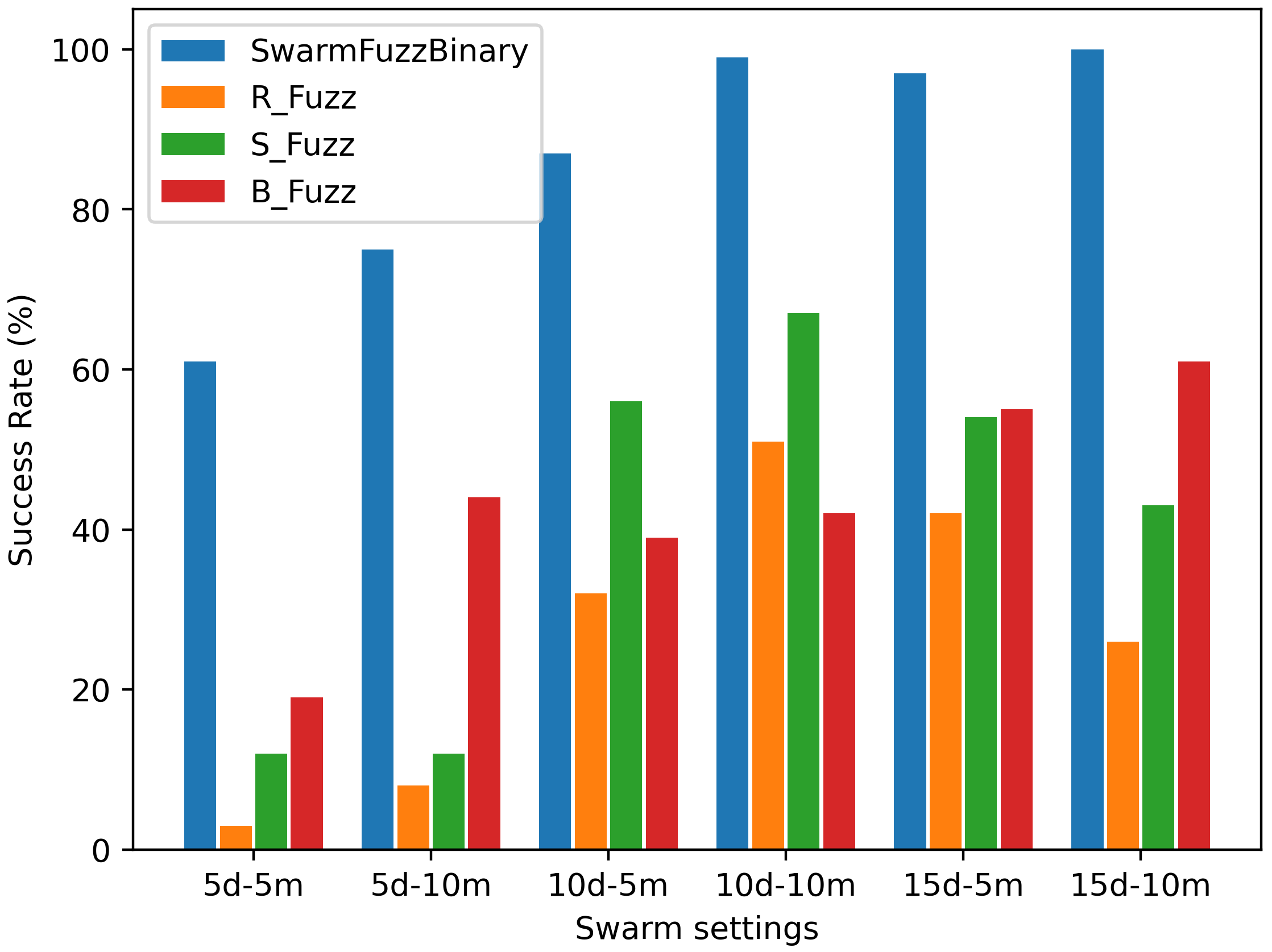} 
        \caption{Success rate (A2)}
        \label{fig:algo2-ablation-suc-A2} 
    \end{subfigure} 
    
    \begin{subfigure}[b]{0.45\linewidth}
        \centering
        \includegraphics[width=0.7\linewidth]{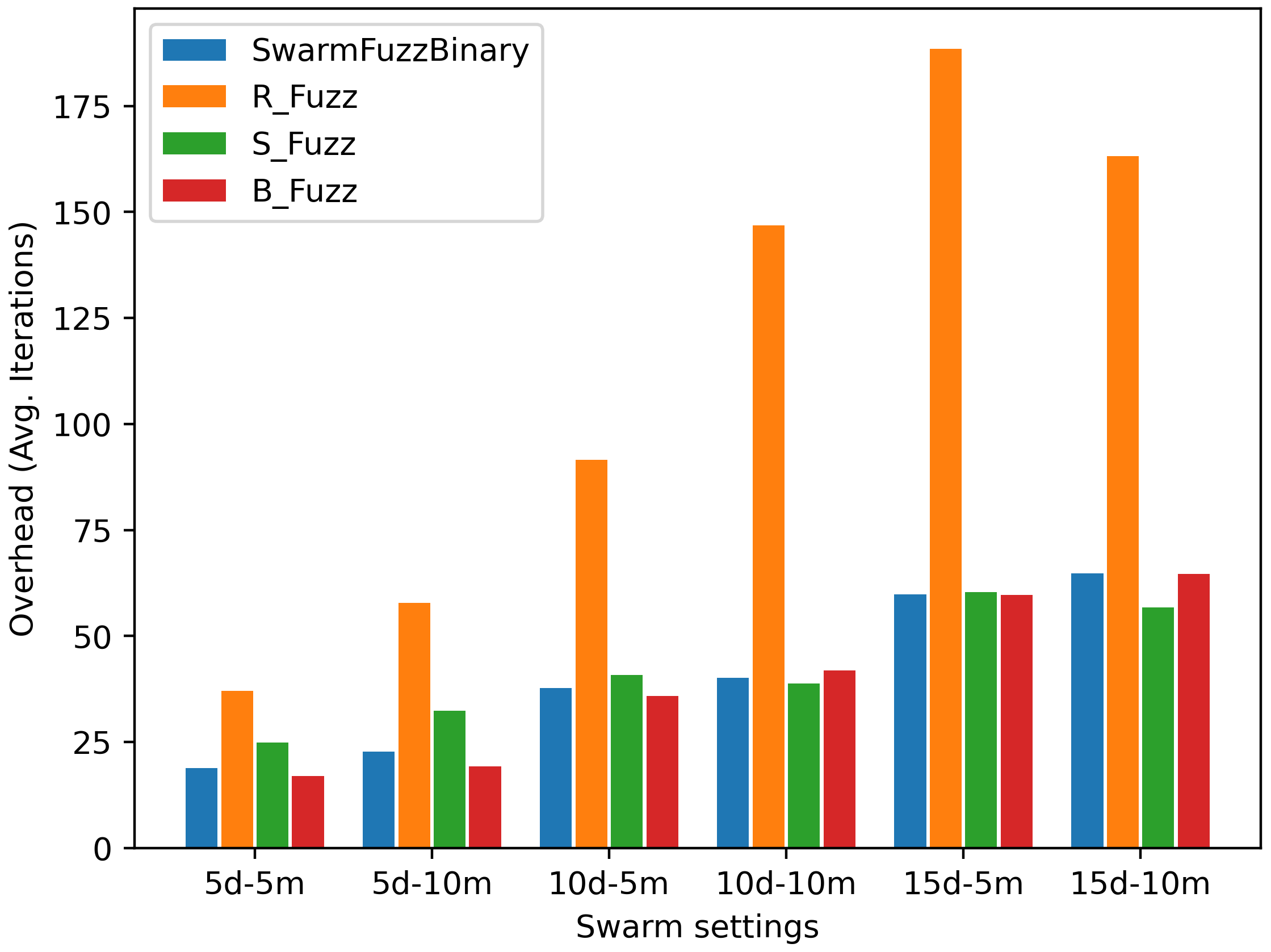} 
        \caption{Runtime (A1)}
        \label{fig:algo2-ablation-over-A1} 
    \end{subfigure}
    \hfill
    \begin{subfigure}[b]{0.45\linewidth}
        \centering
        \includegraphics[width=0.7\linewidth]{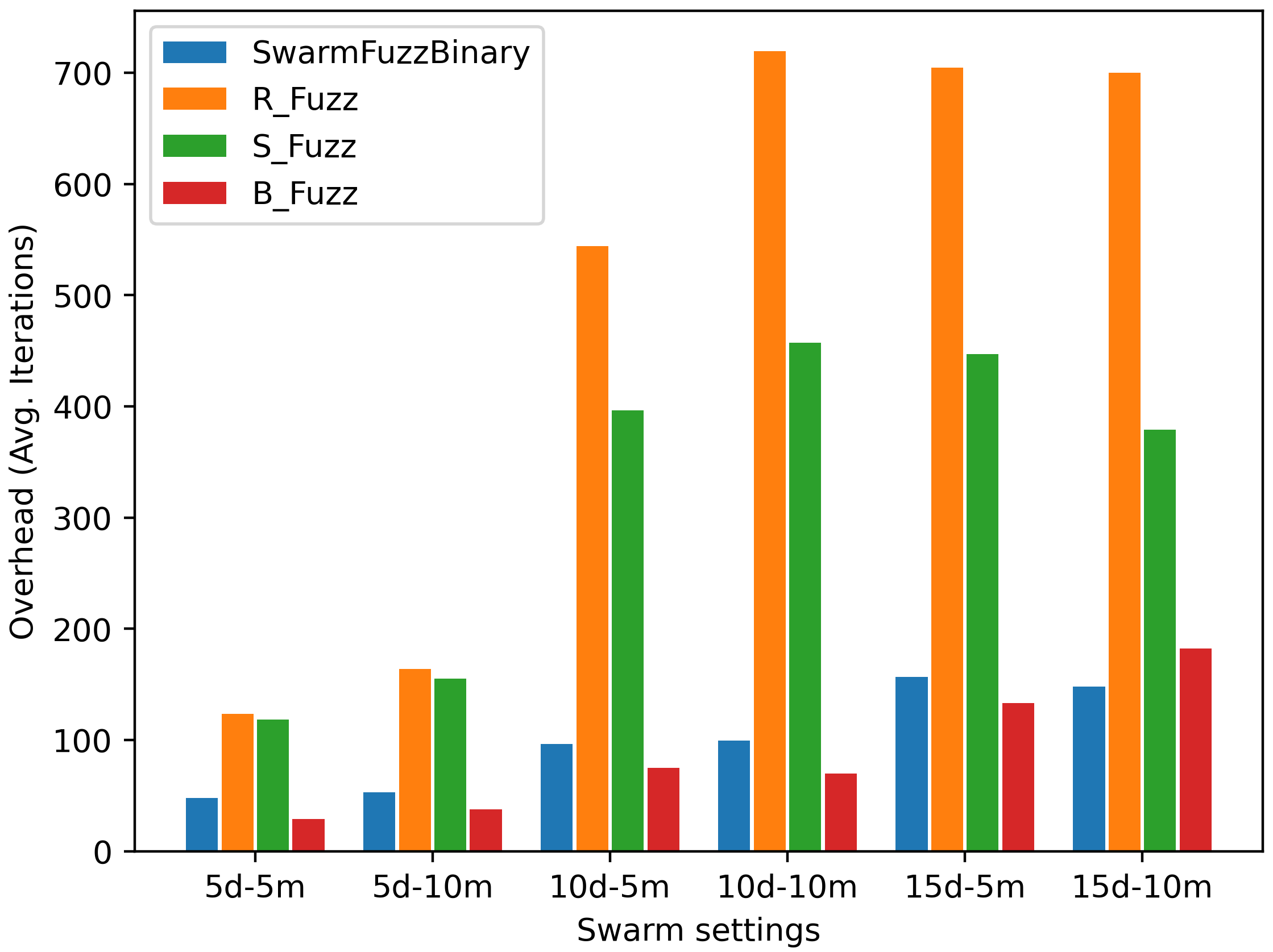} 
        \caption{Runtime (A2)}
        \label{fig:algo2-ablation-over-A2} 
    \end{subfigure} 
    
    \caption{\new{Comparison of fuzzers across A1 and A2 in terms of success rate and runtime.}}
    \label{fig:algo2-ablation}  
\end{figure}

Similar to Section~\ref{sec:ablation1}, R\_Fuzz does not implement either heuristic. B\_Fuzz only implements the binary search to find spoofing parameters, and S\_Fuzz only implements the seed scheduling to choose the drones pairs.
Therefore, we evaluate the efficacy of the seed scheduling by comparing \sysNew with B\_Fuzz, 
and the efficacy of binary search by comparing it with S\_Fuzz.
Figure~\ref{fig:algo2-ablation} shows the results.

\new{We find that the success rate improvement factor of \sysNew ranges from $1.1$X to $20.3$X,  
while the decrease in runtime 
ranges from $0.8$X to $7.2$X.
Specifically, R\_Fuzz and S\_Fuzz fail to find any \atkname in A1. 
On average, compared to S\_Fuzz, the success rate of \sysNew is about $5.4x$ higher, while compared to R\_Fuzz, its runtime is $3.3x$ lower. Thus, the seed scheduling boosts success rate of \sysNew by almost $5.4x$, while the binary search reduces its runtime by about $3.3x$. Compared to the random fuzzer (R\_Fuzz), \sysNew has about $7x$ higher success rate and $5x$ smaller runtime.}

\new{We also find that in A1, S\_Fuzz performs close to \sysNew. This shows that the choice of seeds is critical in finding \atkname. This also suggests that the objective function in A1 has a plateau of global minimums. Therefore, once we identify the promising drone pairs, even searching the GPS parameters randomly could find \atkname.}

\section{Discussion}
\new{We first describe the collision scenarios found by \sysNew on both algorithms. We then compare \sysNew with \sysOld and reflect on the implications of \sysNew. Finally, we present the limitations of both \sysOld and \sysNew.}

\new{\subsection{Collision scenarios}} 
\label{section:vulnerability}
\new{We detail three collision scenarios identified by \sysNew. We first present the scenario difference between two swarm control algorithms, and then describe the underlying reasons causing each collision.}

\new{Overall, the collision scenarios found by \sysNew differ between the two swarm control algorithms. In the Vicsek algorithm (A1), \sysNew causes a collision between a victim drone and an obstacle. However, in the Olfati-Saber algorithm (A2), \sysNew leads to a collision between two victim drones.}

\new{\emph{Case study 1 - Collision with an obstacle (A1).} Drone swarms should avoid colliding with on-path obstacles. Since A1 prioritizes formation maintenance over collision avoidance, A1 even allows a collision in order to adapt to the formation change caused by the target drone.}

\begin{figure}[ht] 
    \centering
    \begin{subfigure}[b]{0.32\linewidth}
        \centering
        \includegraphics[width=0.8\linewidth]{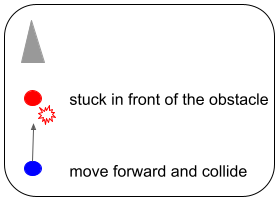} 
        \caption{Trap Scenario (A2)}
        \label{fig:algo2-cs1} 
    \end{subfigure}
    \hfill
    \begin{subfigure}[b]{0.32\linewidth}
        \centering
        \includegraphics[width=0.8\linewidth]{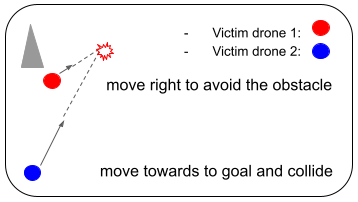} 
        \caption{Trace Intersection Scenario (A2)}
        \label{fig:algo2-cs2} 
    \end{subfigure}
    \hfill
    \begin{subfigure}[b]{0.32\linewidth}
        \centering
        \includegraphics[width=0.8\linewidth]{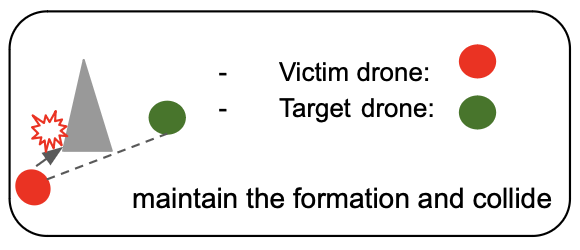} 
        \caption{Collision Scenario (A1)}
        \label{fig:algo1-case} 
    \end{subfigure}

    \caption{Collision scenarios in drone swarm control algorithms. (a) and (b) occur in the Olfati-Saber algorithm (A2), while (c) occurs in the Vicsek algorithm (A1).}
    \label{fig:collision-scenarios}
\end{figure}

\new{To illustrate, Fig.~\ref{fig:algo1-case} depicts the change of the heading direction of the victim drone. After performing GPS spoofing in the target drone, the distance between the victim drone and the target drone increases. To maintain the formation, the Vicsek algorithm generates an attractive velocity between the victim drone and the target drone. Since the Vicsek algorithm prioritizes formation over collision avoidance, the victim drone eventually collides with the obstacle.}

\new{\emph{Case study 2 - Collision in trap scenarios (A2).} In A2, collision avoidance takes priority over formation maintenance. \sysNew{} finds that A2 allows drones to stop to avoid obstacles. This reduces mission efficiency and increases collision risk.}

Fig.~\ref{fig:algo2-cs1} shows how \atkname{} causes two security violations: lack of progress and collisions. Under GPS spoofing, the target drone moves away, triggering an attractive velocity toward victim drone 1 (red). If an obstacle is present, victim drone 1 is pushed toward it but remains trapped due to surrounding drones. Meanwhile, victim drone 2 (yellow) continues moving and collides with the immobilized drone. 
This trap scenario results from a known limitation of potential field methods \cite{pfm}, where robots can get stuck in dead ends, leading to collisions.



\emph{Case study 3 - Collision due to trace Intersections (A2).} Even when a victim drone does not stop to avoid an obstacle, \sysNew{} finds that drones can still collide while maneuvering around it.

\sysNew{} finds that drones can still collide while avoiding obstacles. Fig.~\ref{fig:algo2-cs2} shows this scenario. Under GPS spoofing, victim drone 1 moves toward the target drone to maintain formation and veers right to avoid an obstacle. Meanwhile, victim drone 2, exploring the environment, collides with victim drone 1 due to intersecting paths.


\new{\subsection{Comparison between \sysOld and \sysNew}
Both \sysOld{} and \sysNew{} use the same framework, which is to use seed scheduling to select influential drone pairs, followed by search-based fuzzing to determine spoofing parameters that trigger \atkname{}. However, they differ in each component.}

For the seed scheduling part, \sysOld{} constructs a graph-based model and uses influence analysis to identify key drone pairs in a single iteration, making it efficient. However, this approach does not generalize to different swarm topologies.
In contrast, \sysNew{} observes victim drone behavior during obstacle avoidance to determine influence. This topology-agnostic method adapts to diverse swarm configurations but requires evaluating all drone pairs, increasing runtime.

\new{For the search-based fuzzing part, \sysOld{} models collisions as a convex function and applies gradient descent to find SPV cases, requiring manual parameter tuning.
\sysNew{} converts the function into a monotonic one and uses binary search, eliminating the need for parameter tuning and reducing manual effort.}

\subsection{Implications}



\sysNew finds that swarm missions involving a larger swarm size 
are more vulnerable to attacks exploiting \atkname, since they have low VDOs.
\sysname further finds that  swarm missions with low VDOs are generally more vulnerable.
This suggests that designers should carefully design swarm formation in securing large-size drone swarms.
Therefore, if the VDO is low, then a stricter protection technique against GPS spoofing attacks may be needed. These are future work directions.

\new{From the algorithm aspect, \sysNew finds that although most swarm control algorithms are tuned to balance the conflicting goals under normal conditions, they fail to do so when under sensor attacks (e.g., GPS spoofing). Our work suggests, when tuning the control parameters, the drone swarm developers should take the sensor noise into consideration, to be resilient to \atkname.}

\new{From the sensor aspect, \sysNew finds that even performing small GPS spoofing distance such as $5/10m$ for a short time (e.g., $1s$) can easily cause collisions in the swarm. Our work calls for more accurate and fast GPS spoofing detection, for example, crosschecking the GPS reading with other sensors in the drone.}


\subsection{Limitations}
Our work hasthree limitations.  
First, we only tested \sysOld and \sysNew on two swarm control algorithms.
However, \sysNew does not utilize any knowledge specific to this swarm control algorithm or the mission during fuzzing.
Instead, it only uses general goals designed in the swarm control algorithms and physical properties of collisions, i.e., the convex property of the objective function. 
Therefore, it should also work on other decentralized swarm control algorithms, such as the Adaptive Swarm\cite{Adaptive-swarm} and Sciadro\cite{Sciadro}. These two algorithms are both based on the potential field methods\cite{pfm}, similar to A2 in our evaluation.


Second, in our experiments, we used missions of fixed length, and with a fixed obstacle placement in order to keep the experiment simple.
However, to model other types of missions, for example, \new{with obstacles in different locations}, we only need to change one input - 
the coordinates of the obstacle for collision.
Hence, \sysname should also work on other swarm missions.

\newtcps{Third, we did not consider environmental noise, and hence assumed that the attacker performs constant-value GPS spoofing. In reality, factors such as wind and friction can naturally affect the RV's physical states as well as the GPS spoofing signals. To address this, we can apply domain randomization techniques to introduce environmental noise \cite{env_noise}, such as wind and friction, to improve \sysNew's robustness against real-world disturbances \cite{specguard}.}



\section{Related work}
Physical attack on drones have targeted  GPS\cite{gps-spoofing-1, gps-spoofing-2, gps-spoofing-3, gps-spoofing-4, gps-spoofing-5, gps-spoofing-6, gps-spoofing-7}, 
gyroscope \cite{gyro-spoofing}, accelerometer \cite{acce-spoofing} and optical-flow sensors \cite{optical}.
However, these attacks only focus on single drones, and none of them have considered drone swarms. 
Defense techniques against physical attacks targeting GPS sensors have also been proposed.
Unfortunately, most defense techniques \cite{pidpiper, semperfi} ignore $0-10m$ GPS spoofing distance as it is indistinguishable from the standard GPS offset \cite{gps-offset}, and is sufficient to ensure the safety of a single drone.
\new{Recent work \cite{resilience-signed} uses a statistical anomaly detector for misbehaving agent detection in robotic swarms. However, attackers that add trivial noise to the constant GPS spoofing would evade this and continue to exploit \atkname. }
Thus, these defenses cannot handle attacks exploiting \atkname.

Fuzzing has been used in drones for finding software bugs \cite{rvfuzzer}, policy violation bugs \cite{pgfuzz}, and logic flaws \cite{swarmflawfinder, swarmbug}.
Model checking has been used to evaluate the effects of a single drone's sensor failures \cite{avis}.
SWARMFLAWFINDER \cite{swarmflawfinder} tests the drone swarm's behavior by introducing an external attack drone, which incurs relatively high costs and is easy to detect with intruder detection techniques \cite{intruder}.
Other work finds vulnerabilities in drone swarms by introducing spurious messages in the swarm's  communication~\cite{masquerade}. However, such attacks can be prevented using message encryption \cite{Mavlink}. 
Therefore, prior work cannot find \atkname in drone swarms.

Byzantine consensus design such as blocklist protocol\cite{byzantine_blocklist} and blockchain technology \cite{byzantine-2, byzantine-3} has been applied to ensure the consensus within multi-robot systems\cite{byzantine-4, byzantine-5}.
These approaches focus on the inconsistent behavior exhibited by Byzantine agents. For example, in a drone swarm, a Byzantine drone might transmit different location data to other swarm members, leading to conflicting decisions in swarms. Therefore, these approaches aim to address the consensus within the swarm. 
However, in our attack exploiting \atkname, every swarm member receives the same location of the target drone under GPS spoofing and has already reached the consensus.
Therefore, these Byzantine consensus-based designs are ineffective in our attack.

\section{Conclusion}
This paper introduces Swarm Propagation Vulnerabilities (\atkname{}), a new type of vulnerability in drone swarms that can be exploited via GPS spoofing to cause disruptions such as collisions.
To assess swarm resilience against \atkname{}, we propose two fuzzing tools: \sysOld{} and \sysNew{}. \sysOld{} uses graph centrality analysis to identify influential drone pairs and gradient-guided optimization to find spoofing parameters. Our evaluation on a popular swarm control algorithm shows that \sysOld{} achieves an average success rate of 48.8\%.

However, \sysOld{} struggles with varying swarm topologies. To address this, we introduce \sysNew{}, which replaces graph-based scheduling with observation-based seed scheduling and binary search. Evaluation across two swarm control algorithms shows that \sysNew{} maintains comparable success rates to \sysOld{} while remaining effective across different swarm topologies. Additionally, we find that larger swarms are more vulnerable to \atkname{} for a given spoofing distance.
\newtcps{Our work thus lays the foundations for systematically finding GPS-spoofing vulnerabilities in drone swarm algorithms, and improving their resilience to GPS-spoofing attacks. }


\bibliographystyle{ACM-Reference-Format}
\bibliography{reference}

\end{document}